\documentclass[12pt]{article}
\usepackage{amssymb}
\usepackage{amsfonts}
\usepackage{amsmath}
\usepackage[mathscr]{eucal}
\usepackage{amssymb}
\usepackage{amsthm}
\usepackage{mathrsfs}
\usepackage{bbold}
\usepackage{bm}
\usepackage{graphicx}
\usepackage{caption}
\usepackage[english]{babel}
\usepackage[T2A]{fontenc}
\usepackage[utf8]{inputenc}
\usepackage{cite}
\usepackage{float}
\newcommand{\be}{\begin{equation}}
\newcommand{\ee}{\end{equation}}
\newcommand{\bea}{\begin{eqnarray}}
\newcommand{\eea}{\end{eqnarray}}

\newcommand{\p}[1]{(\ref{#1})}

\def\theequation{\arabic{section}.\arabic{equation}}
\begin{document}
\begin{titlepage}
\vspace*{0.7cm}

\begin{center}
\baselineskip=16pt
{\LARGE\bf $\mathcal{N}{=}\,8$ invariant interaction of dynamical}

\vspace{0.6cm}

{\LARGE\bf  and semidynamical $\mathcal{N}{=}\,4$ multiplets}

%\vskip 0.3cm
\vspace{1.5cm}

{\large\bf Sergey\,Fedoruk,} \quad {\large\bf Evgeny\,Ivanov}
\vspace{0.8cm}

{\it Bogoliubov Laboratory of Theoretical Physics, JINR,\\
Joliot-Curie 6, 141980 Dubna, Moscow region, Russia} \\
\vspace{0.5cm}

{\tt fedoruk@theor.jinr.ru, eivanov@theor.jinr.ru}

\end{center}

\vspace{2cm}

\par
\begin{center}
{\bf Abstract}
\end{center}
We present a new model of $\mathcal{N}{=}\,8$ mechanics with semi-dynamic supermultiplets. The model is constructed as an interaction of $\mathcal{N}{=}\,4$ supermultiplets which carry an implicit $\mathcal{N}{=}\,4$
supersymmetry. The initial field content consists of three dynamical $({\bf 1, 4, 3})$ multiplets: one bosonic and two fermionic. To ensure implicit  $\mathcal{N}{=}\,4$ supersymmetry, we introduce the  superfields
describing three semi-dynamical $({\bf 4, 4, 0})$ multiplets: one fermionic and two bosonic. To avoid the second-order Lagrangian for fermions from the fermionic $({\bf 1, 4, 3})$ multiplets, the conversion of their
velocities into new auxiliary fields is carried out. After conversion, these multiplets turn into semi-dynamical mirror $({\bf 4, 4, 0})$ multiplets without non-canonical terms in the $\mathcal{N}{=}\,8$ Lagrangian at the
component level. The final $\mathcal{N}{=}\,8$ multiplet content is $({\bf 1, 8, 7}) \oplus ({\bf 8, 8, 0})$.  As a first step to the ultimate $\mathcal{N}{=}\,4$ superfield formulation of the model,
we remind a natural description of the standard and mirror $({\bf 4, 4, 0})$ multiplets in the framework of $\mathcal{N}{=}\,4, d{=}\,1$ biharmonic superspace.

\vspace{3.5cm}


\end{titlepage}

\numberwithin{equation}{section}

\section{Introduction}

\quad\, Models of supersymmetric (quantum) mechanics play an important role as a training base for the study of systems with supersymmetry in higher space-time dimensions,
involving the proper supersymmetrizations of gauge, gravitation and cosmological theories. They are also closely related to diverse superextensions of
$d{=}\,1$ integrable systems.

Most advanced and suggestive method to deal with supersymmetric theories is the superfield approach.
Whilst there exists a huge corpus of references on the $d{=}\,1$ superfield description of ${\cal N}{\leq}\, 4$ supersymmetric mechanics models and the related superextended $d{=}\,1$ integrable systems
(see, e.g., reviews \cite{rev1,rev2,rev3}), much less is known about similar approach
to ${\cal N} {>}\, 4$ models, in particular, to ${\cal N}{=}\,8$ ones.
Until now, ${\cal N}{=}\,8$ models (see, e.g.,  \cite{gr,BIKL,ABC,iks}) have been constructed in both ${\cal N}=4$ and ${\cal N}{=}\,8$
superspace approaches\footnote{For a description of ${\cal N}{=}\,8$ supersymmetric systems at the component level,
see, e.g., \cite{bkn,KRT-05,fgh,KT-12,AKT-18,KLS-18,kns}. }.
The superfields used in these models  encompass as a rule ${\cal N}{=}\,8$ supermultiplets of the same type. The latter are dynamical,
that is,  they possess Lagrangians which are of the second order in the time derivatives of the component bosonic fields.
The ${\cal N}{=}\,8$ models involving interactions of different types of ${\cal N}{=}\,4$ multiplets,
with an additional hidden ${\cal N}{=}\,4$ supersymmetry that mixes up these multiplets
and extends the manifest ${\cal N}{=}\,4$ supersymmetry to ${\cal N}{=}\,8$,  were considered, to the best of our knowledge,
only in \cite{F4scm,FI-2019}. Yet, all the involved ${\cal N}{=}\,4$ supermultiplets were of the dynamical type.

On the other hand, a number of models with ${\cal N}{=}\,4$ supersymmetry include, in addition to dynamical supermultiplets,
also semi-dynamical ones.
The bosonic fields of the latter are described by
the $d{=}\,1$ Wess-Zumino (or Chern-Simons) type Lagrangians of the first order in the time derivatives.
The basic goal of the present  work is to construct the first example of ${\cal N}{=}\,8$ supersymmetric models of this sort, with some fields being semi-dynamical.

In \cite{FIL}, an ${\cal N}{=}\,4$ generalization of the $n$-particle rational Calogero system was proposed
(see \cite{superc} for the review).
This ${\cal N}{=}\,4$ Calogero model employs the dynamical $n{\times}n$ matrix $({\bf 1, 4, 3})$ supermultiplet and
$n$ semi-dynamical $({\bf 4, 4, 0})$ supermultiplets. The one-particle ($n{=}\,1$) limit
of the model of ref. \cite{FIL} was considered in \cite{FIL-2009}.
More general form of the kinetic term of the matrix $({\bf 1, 4, 3})$ supermultiplet in the model  of \cite{FIL}
gives rise to the  ${\cal N}{=}\,4$ supersymmetric hyperbolic Calogero-Sutherland model \cite{FIL-2019}.
Superconformal mechanics with $D(2,1,\alpha)$ supersymmetry was constructed in \cite{FIL-2010} as
a generalization of the one-particle system of \cite{FIL-2009}, such that the Lagrangian of the $({\bf 1, 4, 3})$ superfields
is a power function of the latter.

In this paper we construct ${\cal N}{=}\,8$ generalization of ${\cal N}{=}\,4$ system suggested in \cite{FIL-2009}.
We basically use  ${\cal N}{=}\,4, d{=}\,1$
harmonic superspace \cite{IL}, which is a reduction of ${\cal N}{=}\,2, d{=}\,4$ harmonic superspace \cite{GIKOS,HSS}.

Modulo  gauge transformations of the involved superfields, the model of ref. \cite{FIL-2009} is governed  by the cubic action (schematically)
\be\label{act-N4}
\int \mu_{\scriptscriptstyle \mathcal{H}} \,v^2 \ + \ \int \mu_{\scriptscriptstyle \mathcal{A}}^{(-2)} \,\mathcal{V} {Z}^{+ A} {Z}^{+ B} c_{(AB)}\,,
\ee
where $c_{(AB)}$ are some constants. The superfield $v(t,\theta,\bar\theta)$ encapsulates the $({\bf 1, 4, 3})$ supermultiplet,
$\mathcal{V}(t_{\scriptscriptstyle \mathcal{A}},\theta^+,\bar\theta^+,u)$
is its analytic harmonic gauge prepotential and ${Z}^{+ A}(t_{\scriptscriptstyle \mathcal{A}},\theta^+,\bar\theta^+,u)$, $A{=}\,1,2,$
amounts to the  $({\bf 4, 4, 0})$ supermultiplet, which is semi-dynamical in the model with the action \p{act-N4}.
In this paper we find ${\cal N}{=}\,8$ generalization just of the model \p{act-N4}.
We are building this model by making use of ${\cal N}{=}\,4$ superfields carrying an additional implicit ${\cal N}{=}\,4$ supersymmetry.
Similar to \cite{F4scm,FI-2019}, we extend the original ${\cal N}{=}\,4$ superfield content by adding
an odd superfield $\Psi^{+ A}(t_{\scriptscriptstyle \mathcal{A}},\theta^+,\bar\theta^+,u)$ as
a superpartner of the superfield $v(t,\theta,\bar\theta)$ with respect to the implicit ${\cal N}{=}\,4$ supersymmetry.
Analogously, the superfield $\mathcal{Z}^{+ A}(t_{\scriptscriptstyle \mathcal{A}},\theta^+,\bar\theta^+,u)$
is extended to a wider ${\cal N}{=}\,4$ superfield set in which it occupies the same place as the superfield
$\Psi^{+ A}(t_{\scriptscriptstyle \mathcal{A}},\theta^+,\bar\theta^+,u)$ in the first set.
To ensure ${\cal N}{=}\,8$ supersymmetry, we add one more superfield $w(t,\theta,\bar\theta)$ which also has the field content $({\bf 1, 4, 3})$ but it is Grassmann-odd. In order to construct the ${\cal N}{=}\,8$
invariant action it also turns out  necessary to make use of at least two superfields
$\mathcal{Z}_a^{+ A}(t_{\scriptscriptstyle \mathcal{A}},\theta^+,\bar\theta^+,u)$, $a=1,2$ and two superfields
$w_a(t,\theta,\bar\theta)$, $a{=}\,1,2$.
Requiring ${\cal N}{=}\,8$ symmetry for the total extended ${\cal N}{=}\,4$ superfield set, we finally derive the ${\cal N}{=}\,8$ invariant action as a generalization of \eqref{act-N4}.

The plan of the paper is as follows. In Section 2, we present ${\cal N}{=}\,4$ harmonic superfield description of
the multiplets $({\bf 1, 4, 3})$ and  $({\bf 0, 4, 4})$.
Each sort of the supermultiplets involves both even and odd superfields and we describe the implicit ${\cal N}{=}\,4$ supersymmetry
transformations realized on these superfields.
In Section 3, we present the ${\cal N}{=}\,4$ superfield formulation of
the ${\cal N}{=}\,8$ invariant coupling of these supermultiplets.
Section 4 shows that the  superfield model constructed leads to the component Lagrangian in which some fermionic fields enter only through their
first-order time derivative and no such fields without derivatives are present.
After applying the oxidation procedure  of replacing this time derivative by a new auxiliary fermionic field \cite{gr,oxid2}, the Lagrangian
yields ${\cal N}{=}\,8$ invariant model. The resulting ${\cal N}{=}\,8$
invariant model involves on the mass shell one dynamic bosonic field and eight real fermionic fields,
as well as three sets of semi-dynamical bosonic $\mathrm{SU}(2)$-doublet fields.
Some concluding remarks are collected in the last Section 5. In Appendix we demonstrate that the result of the oxidation procedure mentioned above can be reproduced
by using off-shell superfields involving, as elementary components, the auxiliary fermionic fields which imitated the time derivative
of  the  original fermionic fields. This is in agreement with the general proposition of ref. \cite{Iv-11}.

\section{${\cal N}{=}\,4$ superfields used in constructing ${\cal N}{=}\,8$ model}

We will build the ${\cal N}{=}\,8$ model in terms of the ${\cal N}{=}\,4$ superfields defined both  in the usual ${\cal N}{=}\,4$ superspace and
in ${\cal N}{=}\,4$ harmonic superspace \cite{GIKOS,HSS,IL}. In this Section we describe the main features of the objects used throughout the paper.

\subsection{Brief information about ${\cal N}{=}\,4$, $d\,{=}\,1$ harmonic superspace}

\quad\, The powerful approach to constructing ${\cal N}{=}\,4$, $d\,{=}\,1$ supersymmetric models and finding interrelations between them
is ${\cal N}{=}\,4$, $d\,{=}\,1$ harmonic formalism which was proposed in \cite{IL}.
As compared to the description in the usual superspace with the coordinates $z=( t, \theta_i,\bar\theta^i)$, $({\theta_i})^* = \bar\theta^i$
and covariant derivatives
\be
D^i = \frac{\partial}{\partial \theta_i} -
i\bar\theta^i \partial_t\,, \quad \bar D_i = \frac{\partial}{\partial \bar\theta^i} - i\theta_i \partial_t\,, \qquad (D^i)^* = -\bar D_i\,,
\qquad \{D^i, \bar D_k \} = -2i\,\delta^i_k\partial_t\,, \label{defD2}
\ee
the harmonic description involves additional commuting harmonic variables
\be\label{h-2-st}
u^\pm_i\,, \quad (u^+_i)^* = u^-{}^i\,,\qquad u^{+
i}u_i^- =1\,.
\ee

In the harmonic analytic basis
\be\label{h-space}
(z_{\scriptscriptstyle \mathcal{A}},u^\pm_i) = ( t_{\scriptscriptstyle \mathcal{A}}, \theta^\pm, \bar\theta^\pm, u^\pm_i)\,,\qquad t_{\scriptscriptstyle \mathcal{A}} = t +i (\theta^+\bar\theta^- + \theta^-\bar\theta^+), \quad
\theta^\pm =\theta^iu^\pm_i\,, \quad  \bar\theta^\pm = \bar\theta^iu^\pm_i
\ee
half of the ${\cal N}{=}\,4$ covariant spinor derivatives
$D^\pm = u^\pm_i D^i$, $\bar D^\pm = u^\pm_i \bar D^i$  becomes short:
\be
D^+ = \frac{\partial}{\partial \theta{}^-}\,, \qquad \bar D^+ = -\frac{\partial}{\partial \bar\theta{}^-} \,.\label{Short}
\ee
This implies the existence of the harmonic analytic superfields defined on the analytic subspace of the full harmonic superspace:
\be\label{h-2-an}
(\zeta, u) = (t_{\scriptscriptstyle \mathcal{A}}, \theta^+, \bar\theta^+, u^\pm_i)\,, \quad u^{+i}u_i^- =1\,.
\ee
It is closed under ${\cal N}{=}\,4$ supersymmetry and some generalized conjugation, $\widetilde{(t_{\scriptscriptstyle \mathcal{A}})} = t_{\scriptscriptstyle \mathcal{A}}, \,\widetilde{(\theta^+)} = \bar\theta^+,\,\widetilde{(\bar\theta^+)} =  -\theta^+, \,
\widetilde{u^{\pm i}} = u^{\pm}_i, \, \widetilde{u^{\pm }_i} = - u^{\pm i}$.
The integration measure in the harmonic analytic subspace is defined as
$dud\zeta^{(-2)} =dudt_{\scriptscriptstyle \mathcal{A}}d\theta^+d\bar\theta^+$.
An important tool of the formalism is the harmonic derivatives:
\be\label{Dpm}
D^{\pm\pm}=\partial^{\pm\pm}+2i\theta^\pm\bar\theta^\pm\partial_{t_{A}} +\theta^\pm\frac{\partial}{\partial\theta^\mp} +
\bar\theta^\pm\frac{\partial}{\partial\bar\theta^\mp}\,,  \qquad \partial^{\pm\pm} =
u^{\pm}_i\frac{\partial}{\partial u^{\mp}_i}\,.
\ee
The harmonic derivative $D^{++}$ is distinguished in that it commutes with the spinor derivatives \p{Short} and so preserves the analyticity.

The integration measures are defined as
\be\label{measure}
\mu_{\scriptscriptstyle \mathcal{H}}=du\,dt\,d^4\theta=\mu_{\scriptscriptstyle \mathcal{A}}^{(-2)}D^+\bar D^+\,,\qquad
\mu_{\scriptscriptstyle \mathcal{A}}^{(-2)}= du\,d\zeta^{(-2)}\,,\qquad
d\zeta^{(-2)}=dt_{\scriptscriptstyle \mathcal{A}}D^-\bar D^-\,.
\ee

Here we presented only the definitions of the basic notions to be used below.
The full exposition of the harmonic superspace formalism of $d\,{=}\,1$ models can be found in ref. \cite{IL}.

\subsection{${\cal N}{=}\,4$ superfields}

When building the model, we use the following ${\cal N}{=}\,4$ superfields:
\begin{center}
\begin{tabular}{|c|c|}
\hline
& \\
bosonic superfield $v(z)$ & fermionic superfield $\Psi^{iA}(\zeta,u)$ \\
& \\
\hline
& \\
fermionic superfields $w_a(z)$ & bosonic superfields ${\cal Z}_a^{iA}(\zeta,u)$ \\
& \\
\hline
\end{tabular}
\end{center}
Below we describe these superfields in some details.

\subsubsection{Multiplets $({\bf 1, 4, 3})$}

\quad\, The multiplet ${\bf (1, 4, 3)}$ is described by
the ${\cal N}{=}\,4$  even superfield $v(z)$\,, $\widetilde{v} = v$\,,
obeying the constraints \cite{leva}
\be
D^iD_i v = \bar D_i\bar D^i v = 0\,, \quad [D^i, \bar D_i] v = 0\,. \label{Uconstr1}
\ee
The solution of the constraints \p{Uconstr1} is as follows
\begin{equation}  \label{sing-X0-WZ}
v(t,\theta_i,\bar\theta^i)= {\rm v}+ \theta_i\varphi^i +
\bar\theta_i\bar\varphi^i +
i\theta_i\bar\theta_k A^{ik}-{\frac{i}{2}}(\theta)^2\bar\theta_i\dot{\varphi}^i
-{\frac{i}{2}}(\bar\theta)^2\theta_i\dot{\bar\varphi}{}^i +
{\frac{1}{4}}(\theta)^2(\bar\theta)^2 \ddot{\rm v}\,,
\end{equation}
where $(\theta)^2=\theta_k \theta^k$, $(\bar\theta)^2=\bar\theta^k \bar\theta_k$.
The component fields in the expansion \p{sing-X0-WZ}  satisfy the reality conditions
$v^\dag{=}v$, $({\varphi^i})^\dagger=\bar\varphi_i$, $({A^{ik}})^\dagger=A_{ik}=A_{(ik)}$.
In the harmonic superspace the constraints \p{Uconstr1} are rewritten as
\be
D^{++}v =0\,, \qquad D^+D^-v = \bar D^+\bar D^-v = 0\,, \qquad
\left(D^+\bar D^- + \bar D^+D^-\right)v = 0\,.\label{Uconstr2}
\ee
The ${\cal N}{=}\,4$ supersymmetry transformations of the component fields in \p{sing-X0-WZ} are given by
\begin{equation}\label{tr-comp-143}
\begin{array}{c}
\delta_\varepsilon {\rm v}=-\varepsilon_i\varphi^i+ \bar\varepsilon^i\bar\varphi_i\,, \\ [6pt]
\delta_\varepsilon \varphi^i=i \bar\varepsilon^i\dot {\rm v}-i \bar\varepsilon_k A^{ki}  \,,
\qquad
\delta_\varepsilon \bar\varphi_i=-i \varepsilon_i\dot {\rm v} -i \varepsilon^k A_{ki}
\,,\\ [6pt]
\delta_\varepsilon A_{ik}=-2\left( \varepsilon_{(i}\dot\varphi_{k)} +\bar\varepsilon_{(i} \dot{\bar\varphi}_{k)}\right)
\,,
\end{array}
\end{equation}
where $\varepsilon_i$, $\bar\varepsilon^i=(\varepsilon_i)^*$ are odd parameters of the explicit ${\cal N}{=}\,4$ supersymmetry.

As was shown in \cite{2}, the $({\bf 1, 4, 3})$ multiplet can be also described through the real analytic gauge superfield
prepotential ${\cal V}(\zeta, u)$ which is defined up to the abelian gauge transformations
\be\label{VgaugeT}
{\cal V} \qquad\Rightarrow \qquad {\cal V}{\,}' = {\cal V} + D^{++}\Lambda^{--}\,, \quad
\Lambda^{--} = \Lambda^{--}(\zeta, u).
\ee
They allow passing to the Wess-Zumino gauge:
\be
{\cal V}(\zeta, u) = {\rm v}(t_{\scriptscriptstyle \mathcal{A}}) -2 \theta^+\varphi^i(t_{\scriptscriptstyle \mathcal{A}}) u^-_i
-2 \bar\theta^+ \bar\varphi^i(t_{\scriptscriptstyle \mathcal{A}}) u^-_i +
3i \theta^+\bar\theta^+ A^{(ik)}(t_{\scriptscriptstyle \mathcal{A}})u^-_iu^-_k\,. \label{WZV1}
\ee
The superfield $v(z)$ is constructed from the superfield  ${\cal V}(\zeta, u)$  through the transform
\be
v(t, \theta^i, \bar\theta_k)=
\int du\, {\cal V}\left(t +2i\theta^i\bar\theta^k u^+_{(i}u^-_{k)}\,,\, \theta^iu^+_i,
\bar\theta^ku^+_k\,,\, u^\pm_l\right). \label{DefU2}
\ee
The constraints \p{Uconstr1} prove now to be a consequence of the harmonic analyticity constraints
$D^+{\cal V} = \bar D^+{\cal V} = 0$.
The inverse expression of ${\cal V}$ through the superfield $v$ is given by the relation \cite{F4scm}
\be\label{V-v}
{\cal V} = v+D^{++}V^{--} \,,
\ee
where $V^{--}$ is some general harmonic superfield with the transformation law $\delta V^{--}= \Lambda^{--}$
with respect to the gauge transformations \p{VgaugeT}.
In what follows we shall make use of the identity \cite{F4scm}:
\be\label{V-v-eq}
\left(D^+\bar D^- -  \bar D^+D^-\right) v=-2D^+\bar D^+ V^{--} \,.
\ee

In addition to the superfield $v(z)$ we also incorporate
the ${\cal N}{=}\,4$  \textit{odd} superfields $w_a(z)$\,, $a=1,2$\,, $\widetilde{w}_a = -w_a$\,,
obeying the constraints \p{Uconstr1}:
\be\label{Uconstr1-w}
D^iD_i w_a = \bar D_i\bar D^i w_a = 0\,, \quad [D^i, \bar D_i] w_a = 0\,.
\ee
Similarly to \p{sing-X0-WZ}, the constraints \p{Uconstr1-w} have the solution
\begin{equation}  \label{sing-X0-WZ-w}
w_a(t,\theta_i,\bar\theta^i)= {\rm w}_a + \theta_iB_a^i + \bar\theta_i\bar B_{a}^{i} +
\theta_i\bar\theta_k \rho_a^{ik}-{\frac{i}{2}}(\theta)^2\bar\theta_i\dot{B}_{a}^{i}
-{\frac{i}{2}}(\bar\theta)^2\theta_i\dot{\bar B}{}_a^i +
{\frac{1}{4}}(\theta)^2(\bar\theta)^2 \ddot{\rm w}_a\,,
\end{equation}
where the reality conditions for the component fields are as follows,
$(w_a)^\dagger=-w_a$\,, $(B_a^i)^\dagger=\bar B_{ai}$\,, $({\rho_a^{ik}})^\dagger=\rho_{aik}=\rho_{a(ik)}$\,.
In the harmonic superspace the constraints \p{Uconstr1} read
\be\label{Uconstr2-w}
D^{++}w_a =0\,, \qquad D^+D^-w_a = \bar D^+\bar D^-w_a = 0\,, \qquad
\left(D^+\bar D^- + \bar D^+D^-\right)w_a = 0\,.
\ee
The transformation properties of the component fields in the expansion \p{sing-X0-WZ-w} under
${\cal N}{=}\,4$ supersymmetry are given by
\begin{equation}\label{tr-comp-143-w}
\begin{array}{c}
\delta_\varepsilon {\rm w}_a=-\varepsilon_iB_a^i+ \bar\varepsilon^i\bar B_{ai}\,, \\ [6pt]
\delta_\varepsilon B_a^i=i \bar\varepsilon^i\dot {\rm w}_a- \bar\varepsilon_k \rho_a^{ki}  \,,
\qquad
\delta_\varepsilon \bar B_{ai}=-i \varepsilon_i\dot {\rm w}_a - \varepsilon^k \rho_{a\,ki}
\,,\\ [6pt]
\delta_\varepsilon \rho_{a}^{ik}=-2i\left( \varepsilon^{(i}\dot B_a^{k)} +\bar\varepsilon^{(i} \dot{\bar B}_a^{k)}\right)
\,.
\end{array}
\end{equation}

As in \p{VgaugeT}, we can introduce the analytic prepotential superfields ${\cal W}_a(\zeta, u)$
defined up to the proper gauge transformations
\be\label{VgaugeT-w}
{\cal W}_a \qquad\Rightarrow \qquad {\cal W}_a{\,}' = {\cal W}_a + D^{++}\tilde\Lambda_a^{--}\,, \quad
\tilde\Lambda_a^{--} = \tilde\Lambda_a^{--}(\zeta, u)\,.
\ee
In the Wess-Zumino gauge, these superfields read:
\be\label{WZV1-w}
{\cal W}_a(\zeta, u) = {\rm w}_a(t_{\scriptscriptstyle \mathcal{A}}) -2 \theta^+ B_a^i(t_{\scriptscriptstyle \mathcal{A}}) u^-_i
-2 \bar\theta^+ \bar B_a^i(t_{\scriptscriptstyle \mathcal{A}}) u^-_i +
3 \theta^+\bar\theta^+ \rho_a^{(ik)}(t_{\scriptscriptstyle \mathcal{A}})u^-_iu^-_k\,.
\ee
The original superfield $w_a(z)$ is
related to ${\cal W}_a(\zeta, u)$ by the transform
\be\label{DefU2-w}
w_a(t, \theta^i, \bar\theta_k)=
\int du\, {\cal W}_a\left(t +2i\theta^i\bar\theta^k u^+_{(i}u^-_{k)}\,,\, \theta^iu^+_i,
\bar\theta^ku^+_k\,,\, u^\pm_l\right).
\ee
The constraints \p{Uconstr1-w} emerge as a consequence of the harmonic
analyticity of  ${\cal W}_a\,$:
$D^+{\cal W}_a = \bar D^+{\cal W}_a = 0$.
Superfields ${\cal W}_a$ are expressed through superfields $w_a$ as
\be\label{W-w}
{\cal W}_a = w_a+D^{++}W_a^{--} \,,
\ee
where $W_a^{--}$ are some general Grassmann-odd harmonic superfields, such that  $\delta W_a^{--}= \tilde\Lambda_a^{--}$
with respect to the gauge transformation \p{VgaugeT-w}. In the sequel, we shall  use the relations:
\be\label{W-w-eq}
\left(D^+\bar D^- -  \bar D^+D^-\right) w_a=-2D^+\bar D^+ W_a^{--} \,.
\ee

\subsubsection{Multiplets $({\bf 0, 4, 4})$ and $({\bf 4, 4, 0})$}

\quad\, The multiplet $({\bf 0, 4, 4})$ is described by the fermionic
analytic superfield $\Psi^{+ A}$\,,
$\widetilde{(\Psi^{+ A})} = \Psi_A^+$\,, $A=1,2$\,, which satisfies the constraint \cite{IL}:
\be\label{PsiConstr0}
D^{++}\Psi^{+ A} = 0 \,.
\ee
The constraint \p{PsiConstr0} has the general solution
\be\label{PsiConstr}
\Psi^{+ A} =
\psi^{iA}u^+_i + \theta^+ F^A + \bar\theta^+ \bar{F}^A
- 2i\theta^+\bar\theta^+ \dot{\psi}{}^{i A}u^-_i\,,
\ee
where component fields satisfy the reality conditions $({\psi^{iA}})^\dagger=-\psi_{iA}$, $({F^A})^\dagger=\bar{F}_A$.
The doublet index $A\,{=}\,1,2$ is rotated by some Pauli-G\"{u}rsey group $SU(2)_{PG}$ commuting with supersymmetry.
The ${\cal N}{=}\,4$ supersymmetry
transformations of the component fields have the form (see, e.g., \cite{FI-2015}):
\be\label{tr-comp-044}
\delta_\varepsilon\psi^{iA}=-\left( \varepsilon^i F^A+\bar\varepsilon^i \bar F^A\right), \qquad
\delta_\varepsilon F^A=2i\,\bar\varepsilon^k\dot\psi^A_k \,,\qquad
\delta_\varepsilon \bar F_A=2i\,\varepsilon_k\dot{\psi}^{k}_A \,.
\ee
In the central basis, the constraint \p{PsiConstr0} and the analyticity conditions $D^+\Psi^{+ A}{=} \bar D^+\Psi^{+ A}{=}\,0$ imply:
\be\label{ConstrPsi2}
\Psi^{+ A}(z, u) = \Psi^{i A}(z)u^+_i\,,  \qquad
D^{(i} \Psi^{k) A}(z) = \bar D^{(i} \Psi^{k) A}(z) = 0\,,
\ee
where $({\Psi^{iA}})^\dagger=-\Psi_{iA}$.

The multiplets $({\bf 4, 4, 0})$ are described by the bosonic
analytic superfields ${\cal Z}_a^{+ A}$\,,
$\widetilde{({\cal Z}_a^{+ A})} = -{\cal Z}_{aA}^+$\,, $A=1,2$\,, $a=1,2$\,, which satisfy the harmonic constraint \cite{IL}:
\be\label{PsiConstr-z0}
D^{++}{\cal Z}_a^{+ A} = 0\,.
\ee
As a solution to this constraint, the superfields ${\cal Z}_a^{+ A}$ have the following component expansions
\be\label{PsiConstr-z}
{\cal Z}_a^{+ A} =
z_a^{iA}u^+_i + \theta^+ \pi_a^A + \bar\theta^+ \bar{\pi}_a^A
- 2i\theta^+\bar\theta^+ \dot{z}{}_a^{i A}u^-_i\,,
\ee
where $({z_a^{iA}})^\dagger=z_{aiA}$, $({\pi_a^A})^\dagger=\bar{\pi}_{aA}$.
The ${\cal N}{=}\,4$ supersymmetry transformations are realized on the component fields as (see, e.g., \cite{FI-2015}):
\be\label{tr-comp-440}
\delta_\varepsilon z_a^{iA}=-\left( \varepsilon^i \pi_a^A+\bar\varepsilon^i \bar \pi_a^A\right), \qquad
\delta_\varepsilon \pi_a^A=2i\,\bar\varepsilon^k\dot z^A_{ak} \,,\qquad
\delta_\varepsilon \bar \pi_{aA}=2i\,\varepsilon_k\dot{z}^{k}_{aA} \,.
\ee
In the central basis, the constraint \p{PsiConstr-z0} and the analyticity conditions $D^+{\cal Z}_a^{+ A}{=} \bar D^+{\cal Z}_a^{+ A}{=}\,0$
imply
\be\label{ConstrPsi2-z}
{\cal Z}_a^{+ A}(z, u) = {\cal Z}_a^{i A}(z)u^+_i\,,  \qquad
D^{(i} {\cal Z}_a^{k) A}(z) = \bar D^{(i} {\cal Z}_a^{k) A}(z) = 0\,,
\ee
where non-harmonic ${\cal N}=4$ superfields ${\cal Z}_a^{i A}(z)$ are subject to the reality conditions
$({\cal Z}_a^{iA})^\dagger={\cal Z}_{a\,iA}$.

\subsubsection{Implicit ${\cal N}{=}\,4$, $d\,{=}\,1$ supersymmetry}

\quad\, The extra implicit  ${\cal N}{=}\,4$ supersymmetry is realized on the superfields $v(z)$ and $\Psi^{iA}(z)$ by the transformations  \cite{ABC,F4scm}
\be \label{trans-s8}
\delta_\xi v=-\xi_{iA} \Psi^{iA}\,,\qquad  \delta_\xi \Psi^{iA}=\frac12\,\xi_k^A\left( D^i\bar D^k-\bar D^i D^k\right) v\,,
\ee
where $\xi_{iA}=(\xi^{iA})^*$ are fermionic parameters of second ${\cal N}{=}\,4$ supersymmetry.
In terms of the harmonic superfields ${\cal V}(\zeta,u)$ and $\Psi^{+A}(\zeta,u)$ the transformations \p{trans-s8}
take the form \cite{F4scm}
\be \label{trans-s8-v}
\delta_\xi v=\xi^{-A} \Psi^+_{A}-\xi^{+A} \Psi^-_{A}\,,\quad  \delta_\xi {\cal V}=2\xi^{-A} \Psi^+_{A}\,,\qquad
\delta_\xi \Psi^{+A}=D^+\bar D^+\left(\xi^{-A}v + \xi^{+A}V^{--}\right) \,,
\ee
where $\xi^{\pm A}=\xi^{iA}u^\pm_i$. The superfield transformations \p{trans-s8} amount to the following ones for the component fields
\be \label{trans-c8}
\begin{array}{c}
\delta_\xi {\rm v}=-\xi_{iA} \psi^{iA}\,,\qquad
\delta_\xi \varphi^{i}=\xi^{iA}F_A\,,\quad\delta_\xi \bar\varphi_{i}=-\xi_{iA}\bar F^A\,,\qquad
\delta_\xi A_{ik}=2\,\xi_{(iA} \dot\psi^{A}_{k)}\,, \\ [7pt]
\delta_\xi \psi^{iA}=i\xi^{iA}\dot {\rm v} +i \xi^{A}_k A^{ik}\,,\qquad  \delta_\xi F^{A}=-2i\xi_k^A \dot\varphi^k\,,\quad
\delta_\xi \bar F_{A}= 2i\xi^k_A \dot{\bar\varphi}_k\,.
\end{array}
\ee
Thus ${\cal N}=4$ multiplets $({\bf 1, 4, 3})$ and $({\bf 0, 4, 4})$ in  the model under consideration
together constitute  ${\cal N}{=}\,8$ multiplet $({\bf 1, 8, 7})$ \cite{ABC,KRT-05,KT-12,AKT-18}.

Similar implicit  ${\cal N}{=}\,4$ supersymmetry transformations can be defined for ${\cal N}{=}\,4$ superfields $w_a(z)$ and ${\cal Z}_a^{iA}(z)$.
In the conventional superspace these read
\be \label{trans-s8-w}
\delta_\xi w_a=-\xi_{iA} {\cal Z}_a^{iA}\,,\qquad  \delta_\xi {\cal Z}_a^{iA}=\frac12\,\xi_k^A\left( D^i\bar D^k-\bar D^i D^k\right) w_a\,,
\ee
whereas in harmonic space the superfields
${\cal W}_a(\zeta,u)$ and ${\cal Z}_a^{+A}(\zeta,u)$ transform as
\be \label{trans-s8-w-h}
\begin{array}{c}
\delta_\xi w_a=\xi^{-A} {\cal Z}^+_{aA}-\xi^{+A} {\cal Z}^-_{aA}\,,\qquad
\delta_\xi {\cal W}_a=2\xi^{-A} {\cal Z}^+_{aA}\,,\\ [7pt]
\delta_\xi {\cal Z}_a^{+A}=D^+\bar D^+\left(\xi^{-A}w_a + \xi^{+A}W_a^{--}\right) .
\end{array}
\ee
For the component fields these transformations amount to
\be \label{trans-c8-w-comp}
\begin{array}{c}
\delta_\xi {\rm w}_a=-\xi_{iA} z_a^{iA}\,,\qquad
\delta_\xi B_a^{i}=\xi^{iA}\pi_{aA}\,,\quad\delta_\xi \bar B_{ai}=-\xi_{iA}\bar \pi_a^A\,,\qquad
\delta_\xi \rho_{a}^{ik}=2i\,\xi_{A}^{(i} \dot z^{k)A}_{a}\,, \\ [7pt]
\delta_\xi z_a^{iA}=i\xi^{iA}\dot {\rm w}_a + \xi^{A}_k \rho_a^{ik}\,,\qquad  \delta_\xi \pi_a^{A}=-2i\xi_k^A \dot B_a^k\,,\quad
\delta_\xi \bar \pi_{aA}= 2i\xi^k_A \dot{\bar B}_{ak}\,.
\end{array}
\ee

In the next Section we shall construct the interaction of all these superfields which will be invariant under the implicit
${\cal N}{=}\,4$ supersymmetry.

\section{${\cal N}{=}\,8$ invariant coupling}

\quad\, As shown in \cite{ABC,F4scm}, the action
\be \label{free-act}
-\frac12 \int \mu_{\scriptscriptstyle \mathcal{H}} \,v^2 +\frac12 \int \mu_{\scriptscriptstyle \mathcal{A}}^{(-2)}\,\Psi^{+ A}\Psi^{+}_{ A}
\ee
is invariant with respect to the implicit ${\cal N}{=}\,4$ supersymmetry \p{trans-s8}
and describes the free ${\cal N}{=}\,8$ multiplet $({\bf 1, 8, 7})$ in terms of ${\cal N}{=}\,4$ superfields.

Let us build coupling of the multiplets $v$ and $\Psi^{iA}$ to the multiplets $w_a$, $a=1,2$ and ${\cal Z}_a^{iA}$, $a=1,2$.
As the guiding principle we take the requirement of  implicit ${\cal N}{=}\,4$ supersymmetry \p{trans-s8}, \p{trans-s8-w}.
The natural generalization of the second term in the action \p{act-N4} is the action with the analytic Lagrangian
$in^{ab}_{AB}{\cal V}{\cal Z}_a^{+ A}{\cal Z}_{b}^{+B}$, where $n^{ab}_{AB}$ are some constants.
Then, the additional terms needed to ensure the implicit ${\cal N}{=}\,4$ supersymmetry \p{trans-s8}
must have the form $im^{ab}_{AB}{\cal W}_a{\cal Z}_b^{+ A}\Psi^{+B}$,  where $m^{ab}_{AB}$ are some constants.
Thus, we start with  the trial interaction Lagrangian in the form:
\be \label{gen-coupl}
i \int \mu_{\scriptscriptstyle \mathcal{A}}^{(-2)}\Big[n^{ab}_{AB}{\cal V}{\cal Z}_a^{+ A}{\cal Z}_{b}^{+B} \ + \
m^{ab}_{AB}{\cal W}_a{\cal Z}_b^{+ A}\Psi^{+B}\Big].
\ee
Considering only variations $\delta_\xi {\cal V}$, $\delta_\xi {\cal W}_a$ and using \p{trans-s8-v}, \p{trans-s8-w},
we obtain that the corresponding variation of the action \p{gen-coupl} is equal to
\be \label{act-var-gen0}
-2i \int \mu_{\scriptscriptstyle \mathcal{A}}^{(-2)}\xi^-_{C}\Big[n^{ab}_{AB}{\cal Z}_a^{+ A}{\cal Z}_{b}^{+B}\Psi^{+C} \ + \
m^{ab}_{AB}{\cal Z}_a^{+ C}{\cal Z}_{b}^{+A}\Psi^{+B}\Big].
\ee
The quantities $\xi^-_{1}$ and $\xi^-_{2}$ are independent. Therefore,
the requirement of vanishing of \p{act-var-gen0} amounts to the equations
\be \label{eqs-var-gen0}
\begin{array}{rcl}
\left(m^{ab}_{A1}{\cal Z}_a^{+ 2}{\cal Z}_{b}^{+B}\right)\Psi^{+1} \ + \
\left(n^{ab}_{AB}{\cal Z}_a^{+ A}{\cal Z}_{b}^{+B}+m^{ab}_{A2}{\cal Z}_a^{+ 2}{\cal Z}_{b}^{+B}\right)\Psi^{+2}
&=&0\,, \\ [6pt]
\left(m^{ab}_{A2}{\cal Z}_a^{+ 1}{\cal Z}_{b}^{+B}\right)\Psi^{+2} \ + \
\left(n^{ab}_{AB}{\cal Z}_a^{+ A}{\cal Z}_{b}^{+B}+m^{ab}_{A1}{\cal Z}_a^{+ 1}{\cal Z}_{b}^{+B}\right)\Psi^{+1}
&=&0\,.
\end{array}
\ee
Since $\Psi^{+1}$ and $\Psi^{+2}$ are independent, these equations yield the following restrictions on the constants
\be \label{val-const}
m^{ab}_{AB}=2n^{ab}_{AB}=m\epsilon_{ab}\epsilon_{AB}\,,
\ee
where $m$ is a constant. Choosing $m=1/2$, we have \footnote{The superfield action with other choices of the constant $m$ is obtained from action \p{free-coupl}
by the following scale transformation: ${\cal Z}_a^{+ A}\to (2m)^{1/2}{\cal Z}_a^{+ A}$, ${\cal W}_a\to (2m)^{1/2}{\cal W}_a$.}
\be \label{free-coupl}
i \int \mu_{\scriptscriptstyle \mathcal{A}}^{(-2)}\Big[{\cal V}{\cal Z}_1^{+ A}{\cal Z}^{+}_{2 A} \ + \
\left({\cal W}_1{\cal Z}_2^{+ A}-{\cal W}_2{\cal Z}_1^{+ A}\right)\Psi^{+}_{ A}\Big].
\ee

Let us check invariance of \p{free-coupl}
under the implicit ${\cal N}{=}\,4$ supersymmetry \p{trans-s8}.
Considering only variations $\delta_\xi {\cal V}$, $\delta_\xi {\cal W}_a$ and using eqs. \p{trans-s8-v}, \p{trans-s8-w},
we obtain that the corresponding variation of the action \p{free-coupl} is
\be \label{act-var-0}
2i \int \mu_{\scriptscriptstyle \mathcal{A}}^{(-2)}\xi^{-C}\Psi^{+D}{\cal Z}_1^{+ A}{\cal Z}^{+B}_{2}
\left(\epsilon_{AB}\epsilon_{CD}+\epsilon_{AD}\epsilon_{BC}+\epsilon_{AC}\epsilon_{DB}\right),
\ee
and it is identically zero.
In addition, the nullifying of the set of such terms requires the use of
two supermultiplets $w_a$ and two supermultiplets ${\cal Z}_a^{+ A}$ in our construction.

Let us next consider the variation of superfields ${\cal Z}_a^{+ A}$, $\Psi^{+ A}$ in the action \p{free-coupl}.

Consider first the variation of the first term in \p{free-coupl} under the transformation of ${\cal Z}_a^{+ A}$ and
the variation of the second term under that of $\Psi^{+ A}$.
Taking $\delta_\xi {\cal Z}_a^{+A}$ from \p{trans-s8-w-h} and  $\delta_\xi \Psi^{+ A}$ from \p{trans-s8-v}
and using the relation $\mu_{\scriptscriptstyle \mathcal{H}}=\mu_{\scriptscriptstyle \mathcal{A}}^{(-2)}D^+\bar D^+$ (see \p{measure}) for the integration measures, we obtain
\be \label{free-var-l-a}
\begin{array}{l}
\displaystyle
i \int \mu_{\scriptscriptstyle \mathcal{H}}\Big[{\cal V}\left(\xi^{-A}w_1 + \xi^{+A}W_1^{--}\right){\cal Z}^{+}_{2 A} \ - \
{\cal V}\left(\xi^{-A}w_2 + \xi^{+A}W_2^{--}\right){\cal Z}^{+}_{1 A} \\ [6pt]
\qquad\qquad + \
\left({\cal W}_1{\cal Z}_2^{+ A}-{\cal W}_2{\cal Z}_1^{+ A}\right)\left(\xi^{-}_{A}v + \xi^{+}_{A}V^{--}\right)\Big].
\end{array}
\ee
Now we make  the following substitutions in \p{free-var-l-a}: ${\cal V} = v+D^{++}V^{--}$ (eq. \p{V-v})
and ${\cal W}_a = w_a+D^{++}W_a^{--}$ (eq. \p{W-w}).
Half of the terms in the resulting expression contains the superfield ${\cal Z}_1^{+ A}$,
while another half contains  ${\cal Z}_2^{+ A}$.
Those terms in \p{free-var-l-a} which  involve the superfield ${\cal Z}_2^{+ A}$ are collected as:
\be \label{free-var-l-aa}
\begin{array}{l}
\displaystyle
i \int \mu_{\scriptscriptstyle \mathcal{H}}\Big[\left(\xi^{-A}w_1 + \xi^{+A}W_1^{--}\right)\left( v+D^{++}V^{--}\right){\cal Z}^{+}_{2 A} \\ [6pt]
\qquad\qquad + \
\left(\xi^{-A}v + \xi^{+A}V^{--}\right)\left( w_1+D^{++}W_1^{--}\right){\cal Z}^{+}_{2 A}\Big].
\end{array}
\ee
Making in \p{free-var-l-aa} the substitutions $\xi^{+A}=D^{++}\xi^{-A}$ and integrating by parts  with respect to $D^{++}$,
we find that the only surviving term is
$$
-2i \int \mu_{\scriptscriptstyle \mathcal{H}}\,v w_1 \xi^{-A}{\cal Z}^{+}_{2 A}.
$$
It can be rewritten as
\be \label{free-var-l-aaa}
-i \int \mu_{\scriptscriptstyle \mathcal{H}}\,v w_1 \delta_\xi {\cal W}_2
=-i \int \mu_{\scriptscriptstyle \mathcal{H}}\,v w_1 \left(\delta_\xi w_2 + D^{++}\delta_\xi W_2^{--}\right)=-i \int \mu_{\scriptscriptstyle \mathcal{H}}\,v w_1 \delta_\xi w_2 .
\ee
In a similar way, we can show that the terms in \p{free-var-l-a} which contain the superfield ${\cal Z}_1^{+ A}$ are reduced  to
$-i{\displaystyle\int} \mu_{\scriptscriptstyle \mathcal{H}} v\,(\delta_\xi w_1) w_2$.
Thus, the total variation \p{free-var-l-a} proves to be finally equal to
$-i{\displaystyle\int} \mu_{\scriptscriptstyle \mathcal{H}} v\,\delta_\xi (w_1 w_2)$.

It remains to take into account the variation of the second term in  \p{free-coupl}
under the transformations  $\delta_\xi {\cal Z}_a^{+A}$.
Using eq. \p{trans-s8-w-h} for $\delta_\xi {\cal Z}_a^{+A}$,
we find that this variation takes the form:
\be \label{act-var-1}
i \int \mu_{\scriptscriptstyle \mathcal{H}}\Big[
{\cal W}_1\left(\xi^{-A}w_2 + \xi^{+A}W_2^{--}\right)\Psi^{+}_{ A}-
{\cal W}_2\left(\xi^{-A}w_1 + \xi^{+A}W_1^{--}\right)\Psi^{+}_{ A}\Big],
\ee
where we made use of the relation $\mu_{\scriptscriptstyle \mathcal{H}}=\mu_{\scriptscriptstyle \mathcal{A}}^{(-2)}D^+\bar D^+$ (see \p{measure}) for the integration measures.
Substituting the expressions ${\cal W}_a = w_a+D^{++}W_a^{--}$ (eq. \p{W-w}) here, using the conditions $D^{++}\Psi^{+ A} = 0$
(eq. \p{PsiConstr}), and representing $\xi^{+A}=D^{++}\xi^{-A}$,
we find that, modulo a total harmonic derivative in the integrand, the expression \p{act-var-1} is reduced to
\be \label{act-var-2}
-2i{\displaystyle\int} \mu_{\scriptscriptstyle \mathcal{H}} (\xi^{-A}\Psi^+_A)w_1 w_2=-i{\displaystyle\int} \mu_{\scriptscriptstyle \mathcal{H}} (\delta_\xi {\cal V})w_1 w_2
=-i{\displaystyle\int} \mu_{\scriptscriptstyle \mathcal{H}} (\delta_\xi v)w_1 w_2\,.
\ee
In deriving \p{act-var-2} we used that  ${\cal V} = v+D^{++}V^{--}$ (eq. \p{V-v}), $\delta_\xi {\cal V} = \delta_\xi v+D^{++}\delta_\xi V^{--}$
and omitted a total harmonic derivative thanks to the condition $D^{++}w_a=0$ (eq. \p{Uconstr2-w}).

Thus, the total variation of the action \p{free-coupl} under the implicit ${\cal N}=4$ supersymmetry is reduced to
\be \label{act-var-1t}
-i \int \mu_{\scriptscriptstyle \mathcal{H}} \,\delta_\xi (v w_1 w_2).
\ee
As a result, the sum of the action \p{free-coupl} and the action
\be \label{free-coup2}
i \int \mu_{\scriptscriptstyle \mathcal{H}} \,v w_1 w_2
\ee
is invariant with respect to the implicit ${\cal N}{=}\,4$ supersymmetry \p{trans-s8}, \p{trans-s8-w}.

Thus, we have obtained the ${\cal N}{=}\,8$ supersymmetry-invariant action, which is the sum of the actions
\p{free-act}, \p{free-coupl}, \p{free-coup2}
\bea \nonumber
S&=&-\frac12 \int \mu_{\scriptscriptstyle \mathcal{H}} \,v^2 \ + \ \frac12 \int \mu_{\scriptscriptstyle \mathcal{A}}^{(-2)}\,\Psi^{+ A}\Psi^{+}_{ A}
\\ [7pt]
&&
+\,\frac{i}{2}\, \epsilon_{ab}\int \mu_{\scriptscriptstyle \mathcal{H}} \,v w_a w_b
\ + \ \frac{i}{2}\, \epsilon_{ab} \int \mu_{\scriptscriptstyle \mathcal{A}}^{(-2)}\Big[{\cal V}{\cal Z}_a^{+ A}{\cal Z}^{+}_{b A} \ + \
2{\cal W}_a{\cal Z}_b^{+ A}\Psi^{+}_{ A}\Big]
\,. \label{free-full}
\eea

Let us demonstrate that the action \p{free-full} is a generalization of the action \p{act-N4} to the case of two semi-dynamic multiplets ${\cal Z}_a^{+ A}$.
Introducing the superfields $Z^{+ A}$ and $Y^{+ A}$ by the relations
\be \label{z-expans-1}
{\cal Z}_1^{+ A}=Z^{+ A}+i(\sigma_3)^A{}_BY^{+ B}\,,\qquad
{\cal Z}_2^{+ A}=Y^{+ A}-i(\sigma_3)^A{}_BZ^{+ B}\,,
\ee
we obtain
\be \label{act-z-expans-1}
{\cal Z}_1^{+ A}{\cal Z}^{+}_{2 A}\ =\ -iZ^{+ A}Z^{+ B}(\sigma_3)_{AB} -
iY^{+ A}Y^{+ B}(\sigma_3)_{AB}\,,
\ee
where $(\sigma_3)_{AB}=(\sigma_3)_{(AB)}=\epsilon_{AC}(\sigma_3)^C{}_B$.
Thus, in the limit $Y^{+ A}= 0$, $\Psi^{+ A}= 0$, $w_a= 0$
the action \p{free-full} is reduced to the action \p{act-N4} with $c_{AB}=(\sigma_3)_{AB}$.
Of course, when performing the transition ${\cal Z}_a^{+ A}\to (Z^{+ A},Y^{+ A})$ \p{z-expans-1},
the Pauli-G\"{u}rsey $\mathrm{SU}(2)$ symmetry acting on the capital indices $A, B$ gets broken.

The superfield action \p{free-full} can be cast in a more suggestive form:
\bea \nonumber
S&=&-\frac12 \int \mu_{\scriptscriptstyle \mathcal{H}} \left(v -\frac{i}2\, \epsilon_{ab}w_a w_b \right)^2 \ + \
\frac12 \int \mu_{\scriptscriptstyle \mathcal{A}}^{(-2)}\left(\Psi^{+ A}+i\epsilon_{ab}{\cal W}_a{\cal Z}_b^{+ A}\right)
\left(\Psi^{+}_{ A}+i\epsilon_{cd}{\cal W}_c{\cal Z}_{d A}^{+}\right)
\\ [7pt]
&&
+ \, \frac{i}{2}\,  \int \mu_{\scriptscriptstyle \mathcal{A}}^{(-2)}\left({\cal V}-\frac{i}2\, \epsilon_{ab}{\cal W}_a{\cal W}_b\right)
\epsilon_{cd}{\cal Z}_c^{+ A}{\cal Z}^{+}_{d A}
\,. \label{free-full-1}
\eea
The final action \p{free-full-1} contains the scalar composite superfield $\displaystyle v -\frac{i}2\, \epsilon_{ab}w_a w_b$,
the scalar composite analytic superfield $\displaystyle {\cal V}-\frac{i}2\, \epsilon_{ab}{\cal W}_a{\cal W}_b$,
the analytic composite superfields $\Psi^{+ A}+i\epsilon_{ab}{\cal W}_a{\cal Z}_b^{+ A}$,
and the analytic superfields ${\cal Z}_a^{+ A}$.
It is worth to point out  that, although superfield ${\cal V}$ and superfields ${\cal W}_a$ are prepotentials for
the superfields $v$ and $w_a$, respectively, the composite superfield
$\displaystyle {\cal V}-\frac{i}2\, \epsilon_{ab}{\cal W}_a{\cal W}_b$ is by no means a prepotential
for the composite superfield $\displaystyle v -\frac{i}2\, \epsilon_{ab}w_a w_b$.

\section{Component form of ${\cal N}{=}\,8$ action}

\quad\, The superfields entering the action \p{free-full-1} have the following component expansions
\bea\nonumber
&& v -\frac{i}2\,\epsilon_{ab}w_a w_b \ = \ ({\rm v}-\frac{i}2\,\epsilon_{ab}{\rm w}_a{\rm w}_b)
+ \theta_i\left(\varphi^i + i \epsilon_{ab}{\rm w}_a B_b^i\right) +
\bar\theta_i\left(\bar\varphi^i + i \epsilon_{ab}{\rm w}_a \bar B_b^i\right)
\\ [7pt] \nonumber
&&\qquad\qquad
 +\, \frac{i}{4}\,(\theta)^2 \epsilon_{ab}B_a^i B_{bi} -{\frac{i}{4}}(\bar\theta)^2\epsilon_{ab}\bar B_a^i \bar B_{bi}
+ i\theta_i\bar\theta_k \left[ A^{ik}-\epsilon_{ab}\left({\rm w}_a\rho_b^{ik} + B_a^i \bar B_{b}^k \right)\right]
\\ [7pt] \nonumber
&&\qquad\qquad
-\,{\frac{i}{2}}\,(\theta)^2\bar\theta_i\left[\dot{\varphi}^i
+\epsilon_{ab}\left(i{\rm w}_a \dot B_{b}^i+ B_{ak}\rho_b^{ik}\right)\right]
-{\frac{i}{2}}\,(\bar\theta)^2\theta_i\left[\dot{\bar\varphi}{}^i
+\epsilon_{ab}\left(i{\rm w}_a \dot {\bar B}_{b}^i+ \bar B_{ak}\rho_b^{ik}\right)\right]
\nonumber
\\ [7pt]
&&\qquad\qquad
+\,
{\frac{1}{4}}\,(\theta)^2(\bar\theta)^2 \left[\ddot{\rm v}-i \epsilon_{ab}{\rm w}_a\ddot{\rm w}_b
-{\frac{i}{2}}\,\epsilon_{ab}\rho_a^{ik}\rho_{bik}
+\epsilon_{ab}\left(B_{a}^i \dot {\bar B}_{bi} - \dot B_{a}^i {\bar B}_{bi}\right)\right],
\label{v-ww-expr}
\\ [9pt]
 \nonumber
&&{\cal V} -\frac{i}2\,\epsilon_{ab}{\cal W}_a {\cal W}_b \ = \ ({\rm v}-\frac{i}2\,\epsilon_{ab}{\rm w}_a{\rm w}_b)
-2 \theta^+\left(\varphi^- + i \epsilon_{ab}{\rm w}_a B_b^-\right) -
2\bar\theta^+\left(\bar\varphi^- + i \epsilon_{ab}{\rm w}_a \bar B_b^-\right)
\\ [7pt]
&&\qquad\qquad
+\, i\theta^+\bar\theta^+ \left( 3A^{--}-3\epsilon_{ab}{\rm w}_a\rho_b^{--}-4\epsilon_{ab}B_a^- \bar B_{b}^- \right)
,
\label{v-ww-expr-1}
\\ [11pt]
\nonumber
&& \Psi^{+A} +i\epsilon_{ab}{\cal W}_a {\cal Z}_b^{+A} \ = \ (\psi^{+A} +i\epsilon_{ab}w_a z_b^{+A})
\\ [7pt]
\nonumber
&&\qquad\qquad
+\, \theta^+\left[F^A + i \epsilon_{ab}\left({\rm w}_a \pi_b^A+2 z^+_a B^-_b\right)\right]
+ \bar\theta^+\left[\bar F^A + i \epsilon_{ab}\left({\rm w}_a \bar\pi_b^A+2 z^+_a \bar B^-_b\right)\right]
\\ [7pt]
&&\qquad\qquad
-\,2i\,\theta^+\bar\theta^+ \left[\dot\psi^{-A}+
\epsilon_{ab}\left( i{\rm w}_a\dot z_b^{-A}
-\frac32\,\rho^{--}_az_b^{+A} + B_a^- \bar\pi^A_{b} - \bar B_{a}^- \pi_b^A  \right)
\right]
. \label{v-ww-expr-2}
\eea

Inserting \p{v-ww-expr} in the first term of \p{free-full-1}, we see that this term gives rise  to the following component action
$$
\int dt\,\Big(-{\rm v}\ddot{\rm v}+i\ddot{\rm v}{\rm w}_1{\rm w}_2+i{\rm v}\ddot{\rm w}_1{\rm w}_2+
i{\rm v}{\rm w}_1\ddot{\rm w}_2\Big).
$$
Up to a total derivative, this action equals
\be \label{1-Lagr}
\int dt\,\Big(\dot{\rm x}\dot{\rm x} -i\epsilon_{ab}{\rm x} \dot {\rm w}_a\dot {\rm w}_b\Big),
\ee
where
\be \label{x-vww-def}
{\rm x}:={\rm v}-\frac{i}2\,\epsilon_{ab}{\rm w}_a {\rm w}_b\,.
\ee
Thus the model under consideration contains two fermionic fields ${\rm w}_a(t)$, $a=1,2,$ with the second-order Lagrangians for them.

We shall try to bring the action \p{free-full-1} to a form in which it will depend only on $\dot{\rm w}_a(t)$.
By performing the ``oxidation procedure'' \cite{gr,oxid2,bkmo}, in which the quantities $\dot{\rm w}_a(t)$ are replaced by new auxiliary variables,
we get rid of second-order terms in the derivatives of fermionic fields.

In terms of new variables \p{x-vww-def} and
\be \label{n-var-1}
\phi^i := \varphi^i + i \epsilon_{ab}{\rm w}_a B_b^i\,,\qquad
\bar\phi^i := \bar\varphi^i + i \epsilon_{ab}{\rm w}_a \bar B_b^i
\,,\qquad C^{ik} := A^{ik}-\epsilon_{ab}{\rm w}_a\rho_b^{ik}\,,
\ee
\be \label{n-var-2}
\chi^{iA} := \psi^{iA} +i\epsilon_{ab}{\rm w}_a z_b^{iA}\,,\qquad
G^A:=F^A - i \epsilon_{ab}{\rm w}_a \pi_b^A\,,\qquad
\bar G^A := \bar F^A - i \epsilon_{ab}{\rm w}_a \bar\pi_b^A
\,,
\ee
the component off-shell expansions of the superfields \p{v-ww-expr}, \p{v-ww-expr-1}, \p{v-ww-expr-2} are written as
\bea \nonumber
&& v -\frac{i}2\,\epsilon_{ab}w_a w_b \ = \ {\rm x}
+ \theta_i\phi^i  +
\bar\theta_i\bar\phi^i
\\ [7pt] \nonumber
&&\qquad\qquad\qquad
+ \frac{i}{4}\,(\theta)^2 \epsilon_{ab}B_a^i B_{bi} -{\frac{i}{4}}(\bar\theta)^2\epsilon_{ab}\bar B_a^i \bar B_{bi}
+ i\theta_i\bar\theta_k \left[ C^{ik}-\epsilon_{ab} B_a^i \bar B_{b}^k \right]
\\ [7pt] \nonumber
&&\qquad\qquad\qquad
-{\frac{i}{2}}(\theta)^2\bar\theta_i\left[\dot{\phi}^i
+\epsilon_{ab}\left(i B_{a}^i\dot{\rm w}_b + B_{ak}\rho_b^{ik}\right)\right]
-{\frac{i}{2}}(\bar\theta)^2\theta_i\left[\dot{\bar\phi}{}^i
+\epsilon_{ab}\left(i {\bar B}_{a}^i\dot{\rm w}_b+ \bar B_{ak}\rho_b^{ik}\right)\right]
\\ [7pt]
&&\qquad\qquad\qquad
+
{\frac{1}{4}}(\theta)^2(\bar\theta)^2 \left[\ddot{\rm x}+i \epsilon_{ab}\dot{\rm w}_a\dot{\rm w}_b
-{\frac{i}{2}}\,\epsilon_{ab}\rho_a^{ik}\rho_{bik}
+\epsilon_{ab}\left(B_{a}^i \dot {\bar B}_{bi} - \dot B_{a}^i {\bar B}_{bi}\right)\right],
\label{v-ww-expr-r}
\\ [10pt]
&& {\cal V} -\frac{i}2\,\epsilon_{ab}{\cal W}_a {\cal W}_b \ = \ {\rm x}
-2 \theta^+\phi^-  - 2\bar\theta^+\bar\phi^-
+\, i\theta^+\bar\theta^+ \left( 3C^{--}-4\epsilon_{ab}B_a^- \bar B_{b}^- \right)
,
\label{v-ww-expr-1r}
\\ [10pt]
\nonumber
&& \Psi^{+A} +i\epsilon_{ab}{\cal W}_a {\cal Z}_b^{+A} \ = \ \chi^{+A}
+ \theta^+\left(G^A + 2i \epsilon_{ab} z^{+ A}_a B^-_b\right)
+\bar\theta^+\left(\bar G^A + 2i \epsilon_{ab} z^{+ A}_a \bar B^-_b\right)
\\ [7pt]
&&\qquad\qquad\qquad
-\,2i\,\theta^+\bar\theta^+ \left[\dot\chi^{-A}-
\epsilon_{ab}\left( i\dot{\rm w}_a z_b^{-A}
+\frac32\,\rho^{--}_az_b^{+A} - B_a^- \bar\pi^A_{b} + \bar B_{a}^- \pi_b^A  \right)
\right]
. \label{v-ww-expr-2r}
\eea
In the expansions of superfields \p{v-ww-expr-r}, \p{v-ww-expr-1r}, \p{v-ww-expr-2r},
derivatives $\dot{\rm w}_a(t)$ are present, but no fermionic fields ${\rm w}_a(t)$ on their own appear.
Therefore, the component action contains only $\dot{\rm w}_a(t)$ that can be replaced  \cite{gr,oxid2,bkmo} by new fields
\be \label{new-zeta}
\zeta_a(t):=\dot{\rm w}_a(t)\,.
\ee
In terms of these variables the ${\cal N}{=}\,4$ supersymmetry transformation \p{tr-comp-143-w} take the form:
\begin{equation}\label{tr-comp-143-zeta}
\begin{array}{c}
\delta_\varepsilon \zeta_a=-\varepsilon_i\dot B_a^i+ \bar\varepsilon^i\dot {\bar B}_{ai}\,, \\ [6pt]
\delta_\varepsilon B_a^i=i \bar\varepsilon^i\zeta_a- \bar\varepsilon_k \rho_a^{ki}  \,,
\qquad
\delta_\varepsilon \bar B_{ai}=-i \varepsilon_i\zeta_a - \varepsilon^k \rho_{a\,ki}
\,,\\ [6pt]
\delta_\varepsilon \rho_{a}^{ik}=-2i\left( \varepsilon^{(i}\dot B_a^{k)} +\bar\varepsilon^{(i} \dot{\bar B}_a^{k)}\right)
\,.
\end{array}
\end{equation}
After combining $B_a^i$, $\bar B_a^i$, $\zeta_a$, $\rho_{a}^{ik}$ into new fields
\be
f_a{}^{i}_{i'}=(f_a{}^{i}_{i'=1},f_a{}^{i}_{i'=2}):=(B_a^{i},\bar B_a^{i})\,,\qquad
\omega_a{}_{i}{}^k:=i\zeta_a\delta_i^k +\rho_a{}_{i}{}^k\,,
\ee
the transformations \p{tr-comp-143-zeta} are rewritten as
\be\label{tr-comp-440-mir}
\delta f_a^{\, ii'}=\varepsilon^{ki'}\omega_a{}_{k}{}^i\,,\qquad
\delta \omega_a^{\, ik}=-2i\varepsilon^{ij'}\dot f_a{}_{j'}^k\,,
\ee
where the infinitesimal parameters $\varepsilon^{i}$, $\bar\varepsilon^{i}$ are joined
into the $\mathrm{SU}_L(2)\times \mathrm{SU}_R(2)$ bispinor $\varepsilon^{ii'}$:
$\varepsilon^{ii'}=(\varepsilon^{i\,i'=1},\varepsilon^{i\,i'=2})=(\varepsilon^{i},\bar\varepsilon^{i})$.
The indices $i=1,2$ and $i'=1,2$ are acted upon by the $\mathrm{SU}_L(2)$ and $\mathrm{SU}_R(2)$ groups respectively,
which form  the automorphism group $\mathrm{SO}(4)$ of the ${\cal N}{=}\,4$ superalgebra.
For each value of the index $a=1,2$, bosonic $d{=}\,1$ fields $f_a{}^{i}_{i'}$ and fermionic $d=1$ fields $\omega_a^{\, ik}$
are exactly component fields of semi-dynamical $({\bf 4, 4, 0})$ mirror (or twisted) multiplet,
which is described by the superfield $q^{+A'}$ in the bi-harmonic space (see formula (4.7) in \cite{IvNie} and Appendix).
The transformations \p{tr-comp-440-mir} are similar to the transformations \p{tr-comp-440}.
But, the Pauli-G\"{u}rsey group acting on the index $A'$ of mirror multiplet $q^{+A'}$
in the present case was chosen to coincide with the $\mathrm{SU}_L(2)$ group acting on the index $i$ of the original $d=1$ fields.
Thus, after ``oxidation procedure'' two fermionic $({\bf 1, 4, 3})$ multiplets with non-canonical kinetic terms for fermions
transform into two semi-dynamical $({\bf 4, 4, 0})$ mirror multiplets with auxiliary fermions.
The superfield meaning of the mutual conversion of different usual and mirror  ${\cal N}{=}\,4$ multiplets
was clarified in \cite{Iv-11}.

It is important that the ${\cal N}{=}\,4$ supersymmetry transformations of the variables \p{x-vww-def}, \p{n-var-1}, \p{n-var-2}
are expressed only through new variables:
\bea
\delta_\varepsilon {\rm x}&=&-\varepsilon_k\phi^k+\bar\varepsilon^k\bar\phi_k\,, \\ [5pt]
\delta_\varepsilon \phi^i&=&i\bar\varepsilon^i\dot{\rm x}-i\bar\varepsilon_k
\big(C^{ki}+\epsilon_{ab}\bar B_a^k B_b^i \big) +\frac{i}2\,\varepsilon^i\epsilon_{ab}B_{ak}B^k_b\,, \\ [5pt]
\delta_\varepsilon \bar\phi_i&=&-i\varepsilon_i\dot{\rm x}-i \varepsilon^k
\big(C_{ki}-\epsilon_{ab}B_{ak} \bar B_{bi} \big) +\frac{i}2\,\bar\varepsilon_i\epsilon_{ab}\bar B_{ak}\bar B^k_b\,, \\ [5pt]
\delta_\varepsilon C^{ik}&=&-2\big(\varepsilon^{(i}\dot\phi^{k)}+\bar\varepsilon^{(i}\dot{\bar\phi}^{k)}\big)
-2i\epsilon_{ab}\zeta_a\big(\varepsilon^{(i}B_b^{k)}+\bar\varepsilon^{(i}{\bar B}_b^{k)} \big)
+\epsilon_{ab}\big(\varepsilon_j B_a^j+\bar\varepsilon_j{\bar B}_a^j \big)\rho_b^{ik}\,,
\\ [5pt]
\delta_\varepsilon \chi^{iA}&=&-\varepsilon^i G^A-\bar\varepsilon^i \bar G^A
-i\varepsilon_k\epsilon_{ab} B_a^k z^{iA}_b-i\bar\varepsilon_k\epsilon_{ab} \bar B_a^k z^{iA}_b\,,
\\ [5pt]
\delta_\varepsilon G^{A}&=&2i\bar\varepsilon^k\dot\chi^{A}_k+2\bar\varepsilon^k\epsilon_{ab}\zeta_a z^{A}_{bk}
+i\varepsilon_k\epsilon_{ab}B^k_a \pi^{A}_{b} +i\bar\varepsilon_k\epsilon_{ab}\bar B^k_a \pi^{A}_{b}\,,
\\ [5pt]
\delta_\varepsilon \bar G^{A}&=&2i \varepsilon_k\dot\chi^{kA}+2 \varepsilon_k\epsilon_{ab}\zeta_a z^{kA}_{b}
+i\varepsilon_k\epsilon_{ab}B^k_a \bar\pi^{A}_{b} +i\bar\varepsilon_k\epsilon_{ab}\bar B^k_a \bar\pi^{A}_{b}\,.
\eea
The full set of transformations of the new variables also includes the explicit ${\cal N}{=}\,4$ supersymmetry transformations
$\delta_\varepsilon z_a^{iA}$, $\delta_\varepsilon \pi_a^A$, $\delta_\varepsilon \bar \pi_{aA}$ given in
\p{tr-comp-440}.

On the other hand, the implicit ${\cal N}{=}\,4$ supersymmetry transformations \p{trans-c8-w-comp}
involving the variables \p{x-vww-def}, \p{n-var-1}, \p{n-var-2} are of the form
\be
\delta_\xi {\rm x}=-\xi_{iA} \chi^{iA}\,,\qquad
\delta_\xi \chi^{iA}=i\xi^{iA}\dot {\rm x} +i \xi^{A}_k C^{ik}+
\xi_{kB}\epsilon_{ab}z_a^{kB}z_b^{iA}\,,
\ee
\be \label{trans-c8-zeta-comp}
\delta_\xi \zeta_a=-\xi_{iA} \dot z_a^{iA}\,,\qquad
\delta_\xi z_a^{iA}=i\xi^{iA}\zeta_a +i \xi^{A}_k \rho_a^{ik}\,,
\ee
\be
\delta_\xi C^{ik}=2\,\xi_A^{(i} \dot\chi^{k)A} -2i\,\xi_A^{(i}\epsilon_{ab}\zeta_a z_b^{k)A}
+\xi_{jA}\epsilon_{ab}z_a^{jA}\rho_b^{ik}\,,
\ee
\be
\delta_\xi \phi^{i}=\xi^{iA}G_A -i\xi_{kA} \epsilon_{ab}z_a^{kA}B^i_b\,,\qquad
\delta_\xi \bar\phi^{i}=\xi^{iA}\bar G_A-i\xi_{kA} \epsilon_{ab}z_a^{kA}\bar B^i_b\,.
\ee
\be\label{x}
\begin{array}{rcl}
\delta_\xi G^{A}&=&-2i\xi_k^A \dot\phi^k -2\xi_k^A\epsilon_{ab}\zeta_a B^i_b +i\xi_{iB}\epsilon_{ab}z_a^{iB} \pi^A_b\,,\\ [5pt]
\delta_\xi \bar G^{A}&=& -2i\xi_k^A \dot{\bar\phi}^k -2\xi_k^A\epsilon_{ab}\zeta_a \bar B^i_b +i\xi_{iB}\epsilon_{ab}z_a^{iB} \bar \pi^A_b
\end{array}
\ee
and they also contain only new variables. The remaining variations
$\delta_\xi B_a^{i}$, $\delta_\xi \bar B_{ai}$, $\delta_\xi \rho_{a}^{ik}$, $\delta_\xi \pi_a^{A}$, $\delta_\xi \bar \pi_{aA}$
from the complete set of transformations of new variables with respect to the implicit  ${\cal N}{=}\,4$ supersymmetry
are presented in \p{trans-c8-w-comp}.

Thus, the transformations of all new variables \p{tr-comp-143-zeta}-\p{x}, with respect to both explicit and implicit ${\cal N}{=}\,4$ supersymmetries  are expressed only in terms of the new variables.
Therefore, the Lagrangian written in terms of new variables, involving \p{new-zeta}, is invariant under the total ${\cal N}{=}\,8$ supersymmetry.

Using the component expansions \p{PsiConstr-z}, \p{v-ww-expr-1r}, \p{v-ww-expr-2r}, we obtain
\bea \nonumber
&&\frac{\partial}{\partial\bar\theta^+}\frac{\partial}{\partial\theta^+}
\Big[\Big(\Psi^{+A} +i\epsilon_{ab}{\cal W}_a {\cal Z}_b^{+A}\Big)\Big(\Psi_A^{+} +i\epsilon_{cd}{\cal W}_c {\cal Z}_{dA}^{+}\Big)\Big]
=
\\ [7pt]
&& \nonumber\qquad\qquad\qquad
-4i\chi^{+A}\left[\dot\chi^-_A -\epsilon_{ab}\left(i\zeta_az^-_{bA}+\frac32\,\rho^{--}_{a}z^+_{bA}
-B^-_a\bar\pi_{bA}+\bar B^-_a\pi_{bA}\right)\right]
\\ [7pt]
&& \qquad\qquad\qquad \label{ppsi-expr}
+2\left(G^A+2i \epsilon_{ab}z_a^{+A}B^-_b\right)\left(\bar G_A+2i \epsilon_{cd}z_{cA}^{+}\bar B^-_d\right),
\eea
\bea \nonumber
&& \frac{\partial}{\partial\bar\theta^+}\frac{\partial}{\partial\theta^+}
\Big[\Big({\cal V} -\frac{i}2\,\epsilon_{ab}{\cal W}_a {\cal W}_b\Big)\epsilon_{cd}{\cal Z}_c^{+ A}{\cal Z}^{+}_{d A}\Big]
=
\\ [7pt]
&& \nonumber\qquad\qquad\qquad
-2{\rm x}\epsilon_{ab}\left(2iz_a^{+A}\dot z^-_{bA}+\pi_a^A\bar\pi_{bA} \right)
+4\epsilon_{ab}z_a^{+A}\left(\phi^-\bar\pi_{bA} -\bar\phi^-\pi_{bA}\right)
\\ [7pt]
&& \qquad\qquad\qquad \label{v-zz-expr}
+i\epsilon_{ab}z_a^{+A}z_{bA}^{+}\left(3C^{--}-4\epsilon_{cd}B^-_c \bar B^-_d\right).
\eea
Taking into account \p{v-ww-expr-r}, \p{ppsi-expr}, \p{v-zz-expr}
and performing integration over the Grassmann coordinates
[we use $\displaystyle \int \mu_{\scriptscriptstyle \mathcal{H}} (\theta)^2(\bar\theta)^2 K(t)=4 \int dt K(t)\,$,\
$\displaystyle \int \mu_{\scriptscriptstyle \mathcal{A}}^{(-2)} \theta^+\bar\theta^+ N(t_{\scriptscriptstyle \mathcal{A}})= \int dt_{\scriptscriptstyle \mathcal{A}} N(t_{\scriptscriptstyle \mathcal{A}})$], as well as over harmonics,
we derive the off-shell component Lagrangian $L(t)$ corresponding to the action $\displaystyle S=\int dt L(t)$,
defined in  \p{free-full-1}.

This Lagrangian has a somewhat cumbersome form because of the large number of terms present in it:
\bea \nonumber
L &=& \dot{\rm x} \dot{\rm x} + {\rm x}\,\epsilon_{ab}
\left(z_a^{iA}\dot z_{biA}+ \dot B^i_a \bar B_{bi} -B^i_a \dot{\bar B}_{bi}\right)
-i\chi^{iA}\dot\chi_{iA}-i{\bar\phi}^{i}\dot\phi_{i}+i\dot{\bar\phi}^{i}\phi_{i}
\\ [7pt]
\nonumber
&&{} -\,i{\rm x}\, \epsilon_{ab}\zeta_a\zeta_b - \epsilon_{ab}
\left(\bar\phi_{i}B_a^i- \phi_{i}\bar B_a^i-z_{aiA}\chi^{iA}\right)
\zeta_b
\\ [7pt]
\nonumber
&&{} +\,\frac{i}{2}\,{\rm x}\, \epsilon_{ab}\rho_a^{ik}\rho_{bik} +
i \epsilon_{ab}
\left(\bar\phi^{i}B_a^k- \phi^{i}\bar B_a^k-z^i_{aA}\chi^{kA}\right)
\rho_{bik}
\\ [7pt]
\nonumber
&&{} -\,i{\rm x}\, \epsilon_{ab}\pi^A_a\bar\pi_{bA} - 
i\epsilon_{ab}\left(\chi^{iA}B_{ai}- z_a^{iA}\phi_i\right) \bar\pi_{bA}
-i \epsilon_{ab}\pi^A_a\left(\chi^{i}_{A}\bar B_{bi}+ z_{biA}\bar\phi^i\right)
\\ [7pt]
&&{} +\,\frac{1}{2}\,\big(C^{ik}-\epsilon_{ab} B^{(i}_a \bar B^{k)}_{b}\big)\left(C_{ik}-\epsilon_{cd} B_{ci} \bar B_{dk}\right)
- \frac{1}{2}\, \epsilon_{ab} z^{iA}_{a} z_{bA}^{k}C_{ik}
\nonumber
\\ [7pt]
&&{} +\,G^{A}\bar G_{A} +i G^{A}\epsilon_{ab}z^i_{aA}\bar B_{bi}  +i \epsilon_{ab}z^{iA}_{a} B_{bi} \bar G_{A}
\nonumber
\\ [7pt]
&&{} +\,\frac{1}{4}\left(\epsilon_{ab} B^i_a \bar B_{bi}\right)^2
- \frac{1}{4}\, \epsilon_{ab} B^i_a B_{bi}\epsilon_{cd} \bar B^k_c \bar B_{dk}
\nonumber
\\ [7pt]
&&{} -\,\frac{2}{3}\,\epsilon_{ab} z^{(i}_{aA} z_{b}^{k)A}\epsilon_{cd} B_{ci} \bar B_{dk}
+ \frac{4}{3}\,\epsilon_{ab} \epsilon_{cd}z^{(i}_{aA} z_{c}^{k)A} B_{bi} \bar B_{dk}
\,. \label{L-comp-total}
\eea
In \p{L-comp-total} the bosonic fields $C^{ik}$, $G^A$, $\bar G^A$ as well as the
fermionic fields $\rho_a^{ik}$, $\pi_a^A$, $\bar\pi_{aA}$, $\zeta_a$ are auxiliary.
After elimination of these auxiliary fields by their equations of motion, we obtain the following on-shell Lagrangian:
\bea \nonumber
L&=& \dot{\rm x} \dot{\rm x} + {\rm x}\epsilon_{ab}
\left(z_a^{iA}\dot z_{biA}+ \dot B^i_a \bar B_{bi} -B^i_a \dot{\bar B}_{bi}\right)
-i\chi^{iA}\dot\chi_{iA}-i{\bar\phi}^{i}\dot\phi_{i}+i\dot{\bar\phi}^{i}\phi_{i}
\\ [7pt]
\nonumber
&&{} -\,\frac{i}{2{\rm x}}\, \epsilon_{ab}
\left(\bar\phi^{i}B_a^k- \phi^{i}\bar B_a^k-z^i_{aA}\chi^{kA}\right)
\left(\bar\phi_{i}B_{bk}- \phi_{i}\bar B_{bk}-z_{biB}\chi^{B}_k\right) 
\\ [7pt]
\nonumber
&&{} +\frac{i}{{\rm x}}\, \epsilon_{ab}
\left(\chi^{iA}B_{ai}- z_a^{iA}\phi_i\right) 
\left(\chi^{k}_{A}\bar B_{bk}+ z_{bkA}\bar\phi^k\right)
\\ [7pt]
&&{} +\,\frac{1}{4}\left(\epsilon_{ab} B^i_a \bar B_{bi}\right)^2
- \frac{1}{4}\, \epsilon_{ab} B^i_a B_{bi}\epsilon_{cd} \bar B^k_c \bar B_{dk}
- \frac{1}{8}\, \epsilon_{ab} z^{(i}_{aA} z_{b}^{k)A}\epsilon_{cd} z_{ciB} z_{dk}^{B}
\nonumber
\\ [7pt]
&&{} -\,\frac{1}{6}\,\epsilon_{ab} z^{(i}_{aA} z_{b}^{k)A}\epsilon_{cd} B_{ci} \bar B_{dk}
+ \frac{4}{3}\,\epsilon_{ab} \epsilon_{cd}z^{(i}_{aA} z_{c}^{k)A} B_{bi} \bar B_{dk}
- \epsilon_{ab}z^{i}_{aA} B_{bi} \epsilon_{cd} z_{c}^{kA}  \bar B_{dk}
\,.\label{L-comp}
\eea
As follows from the Lagrangian \p{L-comp}, the bosonic variable ${\rm x}$ and fermionic variables $\chi^{ia}, \phi_i$ are dynamical,
while the bosonic variables $z_a^{iA}$, $B^i_a$, $\bar B_a^i$ have the kinetic terms of the first order in $\partial_t$ and so are semi-dynamical.

Thus, under ${\cal N}{=}\,8$ supersymmetrization, we deal with the ${\cal N}{=}\,4$ system which involves on the mass shell
one dynamical bosonic field ${\rm x}$, two semi-dynamical bosonic fields $z_a^{iA}$, as well as additional dynamical fermionic fields $\chi^{iA}$ and
semi-dynamical bosonic fields $B^i_a$, $\bar B_a^i$. It follows from the transformations of implicit ${\cal N}{=}\,4$ supersymmetry given above that the bosonic $({\bf 4,4,0})$ multiplets,
the standard one $(z^{iA}_a, \pi^{A}_a)$ and the mirror one $(B^{i}_a, \bar{B}^i_a, \zeta_a, \rho^{(ik)}_a)$, are transformed through each other and so constitute together the multiplet
$({\bf 8,8,0})$ of ${\cal N}{=}\,8$ supersymmetry \cite{ABC},
while the remaining fields, as was already mentioned, well fit in a kind of $({\bf 1, 8, 7})$ multiplet  \cite{ABC,KRT-05,KT-12,AKT-18}.

\section{Concluding remarks}

\quad\, In this paper we have presented  the ${\cal N}{=}\,8$ supersymmetric model
with dynamical and semi-dynamical $d{=}\,1$ fields.
The initial ``trial'' model \p{free-full-1} was composed from the dynamical  ${\cal N}{=}\,4$ multiplet $({\bf 1, 4, 3})$ (the superfield $v$),
two semi-dynamical bosonic multiplets $({\bf 4, 4, 0})$ (the superfields ${\cal Z}^{+A}_a$)   and
their partners with respect to the implicit ${\cal N}{=}\,4$ supersymmetry (the superfields $\Psi^{+A}$ and $w_a$, respectively).
The latter multiplets have the opposite Grassmann parity as compared to the former ones.

The ${\cal N}{=}\,8$ model constructed describes a system with
the kinetic term of the second order in the ``velocities'' of fermionic fields belonging to $w_a$.
To get rid of this drawback, we carried out the ``oxidation procedure'',
which amounts to replacing the derivatives of fermionic fields with new auxiliary fields \cite{gr,oxid2}.
We have shown that after passing to some suitable new variables such a procedure works perfectly well for our system.
As a result of this procedure, we obtained the new ${\cal N}{=}\,8$ supersymmetric system \p{L-comp}.

On the mass shell the obtained ${\cal N}{=}\,8$
invariant model \p{L-comp} describes one dynamical bosonic field ${\rm x}$ and
eight real fermionic dynamical fields $\phi^{i}$, ${\bar\phi}^{i}$, $\chi^{iA}$,
as well as three sets of semi-dynamical bosonic $\mathrm{SU}(2)$-doublet fields $z_a^{iA}$, $B^i_a$, $\bar B_a^i$.

Surely, the ${\cal N}{=}\,8$ superfield system \p{free-full-1}
and the ${\cal N}{=}\,8$ supersymmetric component system \p{L-comp}
are not equivalent to each other, because the directly applied ``oxidation'' does not preserve the canonical structure of the model.
We have obtained the ${\cal N}{=}\,8$ supersymmetric correct system \p{L-comp} only at the component level.
Rederiving this system at the complete superfield level is the next interesting task.
A clue to this construction might serve the fact that the fields in the action \p{L-comp} naturally fall into a set of
one dynamical ${\cal N}{=}\,8$ multiplet $({\bf 1, 8, 7})$ and one semi-dynamical  ${\cal N}{=}\,8$ multiplet $({\bf 8, 8, 0})$.
When constructing the superfield action, it may happen also necessary to involve some extra auxiliary supermultiplets into the game.
A hint for constructing the self-consistent superfield formulation  is the observation that the transformations \p{tr-comp-143-zeta}
can be identified with the transformations \p{tr-comp-440-mir} of the component fields of two semi-dynamical mirror (or twisted) $({\bf 4, 4, 0})$ multiplets.
In Appendix we demonstrate that such a multiplet has the natural description in the framework of the ${\cal N}{=}\,4, d{=}\,1$ bi-harmonic superfield formalism developed in  \cite{IvNie,Iv-11}.
Capitalizing on this property, we conjecture that the self-consistent superfield formulation of our system can be achieved within such a bi-harmonic stuff \footnote{It is worth noting that the set of fields
of all eventual $({\bf 4, 4, 0})$ multiplets is closed under both manifest and implicit ${\cal N}{=}\,4$ supersymmetries, while the remaining fields are transformed both through themselves and through fields of
$({\bf 4, 4, 0})$ multiplets. This indicates that in the present case we deal with some not fully reducible representation of ${\cal N}{=}\,8$ supersymmetry  and the constraints on ${\cal N}{=}\,4$ superfields belonging to the
$({\bf 1, 8, 7})$ subset should be nonlinear and properly include the $({\bf 4, 4, 0})$ superfields.}.

Another prospective task in the further development of the model constructed is to work out the ${\cal N}{=}\,8$ covariant procedure of gauging
isometries in the systems of this type.
The ${\cal N}{=}\,4$ supersymmetric gauging procedure \cite{2}  proved to be an important tool for construction of ${\cal N}{=}\,4$ supersymmetric generalizations
of integrable many-particle systems of the Calogero type \cite{FIL-2019}.
Being generalized to the ${\cal N}{=}\,8$ case, it would hopefully  provide an opportunity to find out new ${\cal N}{=}\,8$
supersymmetric extensions of these notorious systems.

\bigskip
\section*{Acknowledgements}
The authors would like to thank Armen Nersessian  for useful discussions.
S.A. thanks Sergey Krivonos for useful comments.
This work was supported by the RFBR grant No. 20-52-05008 Arm-a.

\section*{Appendix:
$({\bf 4, 4, 0})$ multiplets in biharmonic superspace}
\def\theequation{A.\arabic{equation}}
\setcounter{equation}0

\quad\, The  automorphism group of ${\cal N}{=}\,4, d=1$ supersymmetry algebra is
$\mathrm{SO}(4)\cong\mathrm{SU}_L(2)\times \mathrm{SU}_L(2)$ group.
Throughout our article, the group $\mathrm{SU}_L(2)$ is implemented explicitly
on the doublet $\mathrm{SU}_L(2)$ indices $i=1,2$.
The harmonics $u_i^\pm$ are associated just with this group.
At the same time, the group $\mathrm{SU}_R(2)$ is implicit.
But there exists the formulation in which both $\mathrm{SU}_L(2)$ and $\mathrm{SU}_R(2)$ symmetries are explicit.
Such a description is achieved in ${\cal N}{=}\,4$ biharmonic superspace \cite{IvNie}
which well suits for describing models where both ${\cal N}{=}\,4$ ordinary and mirror multiplets participate.

In  such a description the odd superspace coordinates $\theta^{i}$, $\bar\theta^{i}$, which are $\mathrm{SU}_L(2)$ doublets, are joined
into  the $\mathrm{SU}_L(2)\times \mathrm{SU}_R(2)$ quartet  $\theta^{ii'}$:
$(\theta^{i},\bar\theta^{i})=(\theta^{i\,i'=1},\theta^{i\,i'=2})$, where
$i=1,2$ and $i'=1,2$ are doublet indices of $\mathrm{SU}_L(2)$ and $\mathrm{SU}_R(2)$, respectively.

In the biharmonic formulation, in addition to the harmonics $u^\pm_i\in \mathrm{SU}_L(2)/\mathrm{U}(1)$ \p{h-2-st},
additional commuting harmonic variables $v^\pm_{i'}\in \mathrm{SU}_R(2)/\mathrm{U}(1)$ are introduced,
%associated with the automorphism group $\mathrm{SU}_R(2)$.
with the defining relations
\be\label{h-2-st-v}
v^\pm_{i'}\,, \quad (v^+_{i'})^* = v^-{}^{i'}\,,\qquad v^{+
{i'}}v_{i'}^- =1\,.
\ee
In the central basis, ${\cal N}{=}\,4$, $d=1$ biharmonic superspace is parametrized by
the coordinates $(t,\theta^{ii'},u^\pm_i,v^\pm_{i'})$.
In this superspace,  we define the harmonic projections of $\theta^{ii'}$ as
\be\label{theta-bi}
\theta^{\pm,\pm} = \theta^{ii'}u^\pm_iv^\pm_{i'}\,, \qquad
\theta^{\pm,\mp} = \theta^{ii'}u^\pm_iv^\mp_{i'}\,.
\ee
Given this,  one of two analytical bases in the biharmonic superspace can be defined: either
with the coordinates
\be\label{an-bas1}
(z_+,u^\pm_i,v^\pm_{i'})\,,\qquad  z_+= ( t_+, \theta^{\pm,\pm}, \theta^{\pm,\mp})\,,\qquad
t_+ = t -i (\theta^{+,+} \theta^{-,-} + \theta^{-,+} \theta^{+,-})
\ee
or with
\be\label{an-bas2}
(z_-,u^\pm_i,v^\pm_{i'})\,,\qquad  z_-= ( t_-, \theta^{\pm,\pm}, \theta^{\pm,\mp})\,,\qquad
t_- = t -i (\theta^{+,+} \theta^{-,-} - \theta^{-,+} \theta^{+,-})\,.
\ee
Note that $t_+$ coincides with the coordinate $t_{\scriptscriptstyle \mathcal{A}}$ introduced in \p{h-space}: $t_+=t_{\scriptscriptstyle \mathcal{A}}$.

In the analytic bases \p{an-bas1} and \p{an-bas2}
half of the ${\cal N}{=}\,4$ covariant spinor derivatives becomes short.
This is a reflection of the fact that the spaces \p{an-bas1} and \p{an-bas2} contain
the ${\cal N}{=}\,4$ invariant subspaces with half the initial Grassmann coordinates.
Namely, analytic superspace parametized by supercoordinates
\be\label{an-bas1a}
(\zeta_+,u^\pm_i,v^\pm_{i'})\,,\qquad  \zeta_+= ( t_+, \theta^{+,+}, \theta^{+,-})
\ee
is closed under the full ${\cal N}{=}\,4$ supersymmetry.
Another analytic superspace,
\be\label{an-bas2a}
(\zeta_-,u^\pm_i,v^\pm_{i'})\,,\qquad  \zeta_-= ( t_-, \theta^{+,+}, \theta^{-,+})\,,
\ee
is also closed.

The ordinary $({\bf 4, 4, 0})$ supermultiplet is described by the superfield
$q^{(+,0)}{}^A(\zeta_+,u,v)$ living in the analytic superspace \p{an-bas1a},
while a mirror multiplet is represented by a superfield $q^{(0,+)}{}^{A'}(\zeta_-,u,v)$ defined on the analytic superspace \p{an-bas2a}.
Here, the indices $A$ and $A'$ are transformed by two Pauli-G\"{u}rsey groups, which are generically different.
These superfields are subject only to the harmonic conditions:
\be\label{harm-conds}
D^{++,0}q^{(+,0)}{}^A=D^{0,++}q^{(+,0)}{}^A=0\,,\qquad  D^{++,0}q^{(0,+)}{}^{A'}= D^{0,++}q^{(0,+)}{}^{A'}=0\,,
\ee
where $D^{++,0}$ and $D^{0,++}$ are the harmonic derivatives
$\partial^{++,0}=u^+_i\partial/\partial u^-_i$ and $\partial^{0,++}=v^+_{i'}\partial/\partial v^-_{i'}$
rewritten in the analytic bases \p{an-bas1} and \p{an-bas2}, respectively (see \p{Dpm}).

Solving the conditions \p{harm-conds} yields the component expansions of the superfields $q^{(+,0)}{}^A$ and $q^{(0,+)}{}^{A'}$.
The ordinary $({\bf 4, 4, 0})$ supermultiplet is described by the superfield
\be\label{ord-440}
q^{(+,0)}{}^A(\zeta_+,u,v)=z^{iA}(t_+)u^+_i + \theta^{+,-}\pi^{i'A}(t_+)v^+_{i'} - \theta^{+,+}\pi^{i'A}(t_+)v^-_{i'}
-2i\theta^{+,+}\theta^{+,-}\partial_{t_+}z^{iA}u^-_i\,,
\ee
while the mirror multiplet is described by the superfield
\be\label{mir-440}
q^{(0,+)}{}^{A'}(\zeta_-,u,v)=f^{i'A'}(t_-)v^+_{i'}+ \theta^{-,+}\omega^{iA'}(t_-)u^+_{i} - \theta^{+,+}\pi^{iA'}(t_-)u^-_{i}
-2i\theta^{+,+}\theta^{-,+}\partial_{t_-}\omega^{i'A'}v^+_{i'}\,.
\ee
Expansion \p{PsiConstr-z} for the superfield ${\cal Z}_a^{+ A}$ at an arbitrary value of $a=1,2$
coincides with the expansion \p{ord-440} for the superfield $q^{(+,0)}{}^A$ after
the following identification of the component fields: $\pi^{i'A}=(\pi^{i'=1\,A},\pi^{i'=2\,A})=(\pi^A ,\bar{\pi}^A)$.
The mirror $({\bf 4, 4, 0})$ multiplets correspond to identifying the index $A'$ in \eqref{mir-440} with $SU(2)_L$ index $j$.
The linear off-shell transformations of the explicit ${\cal N}=4$ supersymmetry on the component fields can easily be obtained from
the standard superfield transformations \footnote{The realizations of implicit ${\cal N}=4$ supersymmetry linearly mixing both $({\bf 4, 4, 0})$ superfields
can also be easily defined \cite{IvNie}.}.

More details on ${\cal N}{=}\,4$ supermultiplets in biharmonic superspace can be found in \cite{IvNie}.


\begin{thebibliography}{99}

\bibitem{rev1}
L.E.\,Gendenshtein, I.V.\,Krive,
{\it Supersymmetry in quantum mechanics},
Sov. Phys. Usp. {\bf 28} (1985) 645.

\bibitem{rev2}
F.\,Cooper, A.\,Khare, U.\,Sukhatme,
{\it Supersymmetry and quantum mechanics},
Phys. Rept. {\bf 251} (1995) 267, {\tt arXiv:hep-th/9405029\,[hep-th]}.

\bibitem{rev3}
P.\,van\,Nieuwenhuizen,
{\it Supersymmetry, supergravity, superspace and BRST symmetry in a simple model},
Proc. Symp. Pure Math. {\bf 73} (2005) 381, {\tt hep-th/0408179\,[hep-th]}.

\bibitem{gr}
S.J.\,Gates,\,Jr., L.\,Rana,
{\it Ultramultiplets: A new representation of rigid 2-d, ${\cal N}{=}\,8$ supersymmetry},
Phys. Lett. {\bf B342} (1995) 132, {\tt arXiv:hep-th/9410150\,[hep-th]}.

\bibitem{BIKL}
S.\,Bellucci, E.\,Ivanov, S.\,Krivonos, O.\,Lechtenfeld,
{\it ${\cal N}{=}\,8$ superconformal mechanics},
Nucl. Phys. B {\bf 684} (2004) 321, {\tt arXiv:hep-th/0312322}.


\bibitem{ABC}
S.\,Bellucci, E.\,Ivanov, S.\,Krivonos, O.\,Lechtenfeld,
{ABC of $\mathcal{N}{=}\,8$, $d{=}\,1$ supermultiplets},
Nucl. Phys. B {\bf 699} (2004) 226, {\tt arXiv:hep-th/0406015}.

\bibitem{iks}
E.\,Ivanov, O.\,Lechtenfeld, A.\,Sutulin,
{\it Hierarchy of ${\cal N}{=}\,8$ mechanics models},
Nucl. Phys. B {\bf 790} (2008) 493, {\tt arXiv:0705.3064\,[hep-th]}.

\bibitem{bkn}
S.\,Bellucci, S.\,Krivonos, A.\,Nersessian,
{\it ${\cal N}{=}\,8$ supersymmetric mechanics on special K\"{a}hler manifolds},
Phys. Lett. B {\bf 605} (2005) 181, {\tt arXiv:hep-th/0410029}.

\bibitem{KRT-05}
Z.\,Kuznetsova, M.\,Rojas, F.\,Toppan,
{\it Classification of irreps and invariants of the $\mathcal{N}$-extended supersymmetric quantum mechanics},
JHEP {\bf 03} (2006) 098, {\tt arXiv:hep-th/0511274}.

\bibitem{fgh}
M.G.\,Faux, S.J.\,Gates,\,Jr., T.\,Hubsch,
{\it Effective symmetries of the minimal supermultiplet of ${\cal N}{=}\,8$ extended worldline supersymmetry},
J. Phys. A {\bf 42} (2009) 415206, {\tt arXiv:0904.4719\,[hep-th]}.

\bibitem{KT-12}
S.\,Khodaee, F.\,Toppan,
{\it Critical scaling dimension of D-module representations of $\mathcal{N}=4,7,8$ Superconformal Algebras and constraints on Superconformal Mechanics},
J. Math. Phys. {\bf 53} (2012) 103518, {\tt arXiv:1208.3612\,[hep-th]}.

\bibitem{AKT-18}
N.\,Aizawa, Z.\,Kuznetsova, F.\,Toppan,
{\it The quasi-nonassociative exceptional F(4) deformed quantum oscillator},
J. Math. Phys. {\bf 59} (2018) 022101, {\tt arXiv:1711.02923\,[math-ph]}.


\bibitem{KLS-18}
S.\,Krivonos, O.\,Lechtenfeld, A.\,Sutulin,
{\it  $\mathcal{N}$-extended supersymmetric Calogero models},
Phys. Lett. B {\bf 784} (2018) 137,
{\tt arXiv:1804.10825 [hep-th]}.

\bibitem{kns}
S.\,Krivonos, A.\,Nersessian, H.\,Shmavonyan,
{\it Geometry and integrability in ${\cal N}{=}\,8$ supersymmetric mechanics},
Phys. Rev. D {\bf 101} (2020) 045002, {\tt arXiv:1908.06490\,[hep-th]}.



\bibitem{F4scm}
F.\,Delduc, E.\,Ivanov,
{\it New model of ${\cal N}{=}\,8$ superconformal mechanics},
Phys. Lett. B {\bf 654} (2007) 200, {\tt arXiv:0706.2472\,[hep-th]}.


\bibitem{FI-2019}
S.\,Fedoruk, E.\,Ivanov,
{\it Multiparticle ${\cal N}{=}\,8$ mechanics with $F(4)$ superconformal symmetry},
Nucl. Phys. B {\bf 938} (2019) 714, {\tt arXiv:1810.13366\,[hep-th]}.


\bibitem{FIL}
S.\,Fedoruk, E.\,Ivanov, O.\,Lechtenfeld,
{\it Supersymmetric Calogero models by gauging},
Phys. Rev. D {\bf 79} (2009) 105015,
{\tt arXiv:0812.4276\,[hep-th]}.

\bibitem{superc}
S.\,Fedoruk, E.\,Ivanov, O.\,Lechtenfeld,
{\it Superconformal mechanics},
J. Phys. A  {\bf 45} (2012) 173001,
{\tt arXiv:1112.1947\,[hep-th]}.


\bibitem{FIL-2009}
S.\,Fedoruk, E.\,Ivanov, O.\,Lechtenfeld,
{\it $OSp(4|2)$ superconformal mechanics},
JHEP {\bf 08} (2009) 081, {\tt arXiv:0905.4951\,[hep-th]}.

\bibitem{FIL-2019}
S.\,Fedoruk, E.\,Ivanov, O.\,Lechtenfeld,
{\it Supersymmetric hyperbolic Calogero-Sutherland models by gauging},
Nucl. Phys. B {\bf 944} (2019) 114633, {\tt arXiv:1902.08023\,[hep-th]}.


\bibitem{FIL-2010}
S.\,Fedoruk, E.\,Ivanov, O.\,Lechtenfeld,
{\it New $D(2,1,\alpha)$  mechanics with spin variables},
JHEP {\bf 04} (2010) 129, {\tt arXiv:0912.3508\,[hep-th]}.

\bibitem{IL}
E.\,Ivanov, O.\,Lechtenfeld,
{\it ${\mathcal{N}}{=}\,4$ supersymmetric mechanics in harmonic superspace},
JHEP {\bf 0309} (2003) 073, {\tt arXiv:hep-th/0307111}.

\bibitem{GIKOS}
A.S.\,Galperin, E.A.\,Ivanov, S.\,Kalitzin, V.I.\,Ogievetsky, E.S.\,Sokatchev,
{\it Unconstrained ${\cal N}{=}\,2$ matter, Yang-Mills and supergravity
theories in harmonic superspace}, Class. Quant. Grav.
{\bf 1} (1984) 469.

\bibitem{HSS}
A.S.\,Galperin, E.A.\,Ivanov, V.I.\,Ogievetsky and E.S.\,Sokatchev,
{\it Harmonic Superspace}, Cambridge University Press 2001, 306 p.

\bibitem{oxid2}
A.\, Pashnev, F.\,Toppan,
{\it On the classification of N extended supersymmetric quantum mechanical systems},
J. Math. Phys. {\bf 42} (2001) 5257-5271, {\tt arXiv:hep-th/0010135}.

\bibitem{Iv-11}
E.\,Ivanov,
{\it Harmonic Superfields in $\mathcal{N}{=}\,4$ supersymmetric quantum mechanics},
SIGMA {\bf 7} (2011) 015, {\tt arXiv:1102.2288\,[hep-th]}.

\bibitem{leva}
E.\,Ivanov, S.\,Krivonos, V.\,Leviant,
{\it Geometric superfield approach to superconformal mechanics},
J.\,Phys.\,A: Math. Gen. {\bf 22} (1989) 4201.

\bibitem{2}
F.\,Delduc, E.\,Ivanov, {\it Gauging ${\cal N}{=}\,4$ supersymmetric mechanics}, Nucl. Phys. B {\bf 753} (2006) 211,
{\tt arXiv:hep-th/0605211};
{\it Gauging ${\cal N}{=}\,4$ supersymmetric mechanics II: (1,4,3) models from the (4,4,0) ones},
Nucl. Phys. B {\bf 770} (2007) 179, {\tt arXiv:hep-th/0611247}.

\bibitem{FI-2015}
S.\,Fedoruk, E.\,Ivanov,
{\it New realizations of the supergroup D(2,1;$\alpha$) in $\mathcal{N}{=}\,4$ superconformal mechanics},
JHEP {\bf 1510} (2015) 087, {\tt arXiv:1507.08584\,[hep-th]}.


\bibitem{bkmo}
S.\,Bellucci, S.\,Krivonos, A.\,Marrani, E.\,Orazi,
{\it `Root' action for ${\cal N}{=}\,4$ supersymmetric mechanics theories},
Phys. Rev. {\bf D73} (2006) 025011, {\tt arXiv:hep-th/0511249}.

\bibitem{IvNie}
E.\,Ivanov, J.\,Niederle,
{\it Bi-harmonic superspace for $\mathcal{N}{=}\,4$ mechanics},
Phys. Rev. D {\bf 80} (2009) 065027,  {\tt arXiv:0905.3770\,[hep-th]}.



\end{thebibliography}
\end{document}